\documentclass[aps,pre,twocolumn,groupedaddress,showpacs,floatfix,suprscriptaddress]{revtex4-1}
\usepackage[utf8]{inputenc}
\usepackage[T1]{fontenc}
\usepackage{graphicx}
\usepackage{amssymb}
\usepackage{nicefrac}
\usepackage{xcolor}
\usepackage{here}

\usepackage{times}

\usepackage[colorinlistoftodos,prependcaption,textsize=small]{todonotes}

\usepackage{amsmath}
\DeclareMathOperator{\sign}{sign}

\usepackage{hyperref}

\hypersetup{
  colorlinks=true,
citecolor=black,
linkcolor=black,
urlcolor=black
  }

\newcommand{\pd}[3]{\frac{ \partial^{ #1 } #2 }{ \partial #3^{ #1 } }}

\begin{document}
\preprint{AIP/123-QED}

\title{Stationary states for underdamped anharmonic oscillators driven by Cauchy noise}

\author{Karol Capa{\l}a}
\email{karol@th.if.uj.edu.pl} \affiliation{Marian Smoluchowski
Institute of Physics, and Mark Kac Center for Complex Systems
Research, Jagiellonian University, ul. St. {\L}ojasiewicza 11,
30--348 Krak\'ow, Poland}

\author{Bart{\l}omiej Dybiec}
\email{bartek@th.if.uj.edu.pl} \affiliation{Marian Smoluchowski
Institute of Physics, and Mark Kac Center for Complex Systems
Research, Jagiellonian University, ul. St. {\L}ojasiewicza 11,
30--348 Krak\'ow, Poland}

\date{\today}

\begin{abstract}

Using numerical methods,
we have studied stationary states in the underdamped anharmonic stochastic oscillators driven by Cauchy noise.
Shape of stationary states depend both on the potential type and the damping.
If the damping is strong enough, for potential wells which in the overdamped regime produce multimodal stationary states, stationary states in the underdamped regime can be multimodal with the same number of modes like in the overdamped regime.
For the parabolic potential, the stationary density is always unimodal and it is given by the two dimensional $\alpha$-stable density.
For the mixture of quartic and parabolic single-well potentials the stationary density can be bimodal. Nevertheless,  the parabolic addition, which is strong enough, can destroy bimodlity of the stationary state.

\end{abstract}


\maketitle

%%%%%%%%%%%%%%%%%%%%%%%%%%%%%%%%%%%%%%%%%%%%%%%%%%%%%%%%%%%%%%%%%%%%%%%%
\textbf{The increasing number of observations shows that fluctuations in complex systems do not need to follow the Gaussian distribution but display power-law behavior. T
he non-equilibrium fluctuations can be approximated and modeled by $\alpha$-stable (L\'evy) noise. Properties of dynamical systems driven by L\'evy noise are significantly different from their Gaussian white noise driven counterparts.
This is especially well visible in noise induced effects but also in stationary states.
Here, we study the archetypal models of anharmonic, inertial stochastic oscillators driven by the L\'evy noise.
Therefore, within the current manuscript  we extend understanding of uderdamped, L\'evy noise driven systems.
We demonstrate that stationary states strongly deviate from the Boltzmann-Gibbs distribution as in single-well potentials multimodal stationary densities can be observed. Moreover, contrary to the Gaussian driving, stationary densities depend on the damping. Finally, we show under which conditions stationary states in singe-well potentials can be multimodal.}

\section{Introduction}
The overdamped Langevin eqution
\begin{equation}
     \dot{x}(t) = -V'(x) + \zeta (t).
    \label{eq:langevin}
\end{equation}
is the archetypal equation in the theory of stochastic systems.
It describes the evolution of the position $x(t)$ of the overdamped, noise driven particle moving in the potential $V(x)$.
The potential $V(x)$ produces the deterministic force $f(x)=-V'(x)$, while $\zeta(t)$ stands for the noise, which approximates (random) interactions of the observed particle with its environment.
In the simplest realms it is assumed that the noise $\zeta(t)$ is white and Gaussian \cite{horsthemke1984,gardiner1985stochastic}.
For such a noise  $\langle \zeta(t) \rangle=0$ and $\langle \zeta(t)\zeta(s) \rangle = \sigma ^2 \delta(t-s)$.

Numerous extensions of Eq.~(\ref{eq:langevin}) to time dependent forces $f(x,t)$ account for description of various noise induced effects: stochastic resonance \cite{gammaitoni2009}, resonant activation \cite{doering1992}, ratcheting effect \cite{magnasco1993,reimann2002,li2017transports}, to name a few.
Eq.~(\ref{eq:langevin}) also underlines description of stationary states in noisy systems, which constitute the main topic of current research.
If $\zeta(t)$ is the Gaussian white noise,  a stationary state exists for any potential well such that $V(x) \to \infty$ as $|x|\to\infty$.  It is given by the Boltzmann-Gibbs distribution, i.e. $P(x) \propto \exp[-V(x)/\sigma^2]$, see Refs.~\onlinecite{horsthemke1984,reichl1998}.

Noise in Eq.~(\ref{eq:langevin}) does not need to be Gaussian.
A natural generalization of the Gaussian white noise is provided by the L\'evy noise. $\alpha$-stable, L\'evy type  noise is a non-equilibrium noise which is the formal time derivative of the $\alpha$-stable process $L(t)$, see Ref.~\onlinecite{janicki1994}, whose probability density follow an  $\alpha$-stable density \cite{janicki1994,janicki1996} with the scale parameter which grows in time. The characteristic function  (Fourier transform) $\phi(k)=\langle \exp[i k L(t)] \rangle$ of symmetric L\'evy process is given by
\begin{equation}
 \phi(k)=\exp\left[ - \sigma^\alpha(t) |k|^\alpha \right].
 \label{eq:fcharakt}
\end{equation}
Symmetric $\alpha$-stable densities are unimodal probability densities with power-law tails.
The stability index $\alpha$ ($0<\alpha \leqslant 2$) describes the tails asymptotic which for $\alpha<2$ is of  $ |x|^{-(\alpha+1)}$ type. The scale parameter $\sigma$ controls the distribution width.
For the L\'evy motion it  grows in time as $\sigma(t) = \sigma_0 t^{1/\alpha}$, where $\sigma_0$ is the scale parameter characterizing the strength of L\'evy noise $\zeta(t)$.
More precisely, increments of the L\'evy process $\Delta L=L(t+\Delta t)-L(t)$ are distributed according to the $\alpha$-stable density with the characteristic function $\exp[\Delta t \sigma_0^\alpha  |k|^\alpha]$.
For $\alpha<2$, the variance of an $\alpha$-stable density is infinite, thus the distribution width can be defined by the interquantile width or fractional moments only \cite{samorodnitsky1994,janicki1994}.
For $\alpha=2$, the L\'evy noise is equivalent to the Gaussian white noise.
In the most general scenario, not considered here, the L\'evy noise can be asymmetric and shifted \cite{samorodnitsky1994,janicki1994}.
Non-gaussian, heavy tailed fluctuations have been observed in numerous experimental setups \cite{bouchaud1990,bouchaud1991,klages2008,solomon1993,barthelemy2008,cabrera2004} and used in description of multiple phenomena \cite{mercadier2009levyflights,cohen1990,barkai2014}, for a review see Ref.~\onlinecite{dubkov2008}.
Moreover, in the last two decades theory of systems driven by the L\'evy noise has been significantly advanced \cite{metzler2000,chechkin2000linear,barkai2001,anh2003,brockmann2002,chechkin2006,jespersen1999,yanovsky2000,schertzer2001,eliazar2003,sokolov2004b,garbaczewski2009,garbaczewski2010}.

The problem of stationary states in overdamped systems driven by $\alpha$-stable noises has been studied for the long time.
It is well known that stationary states do not exist for all type of potential wells.
Moreover, if they exist, they are not of the Boltzmann-Gibbs type \cite{eliazar2003}.
For single-well potentials of $V(x)=|x|^\nu/\nu$ type ($\nu>0$) stationary states exist for sufficiently large $\nu$.
Surprisingly, the limiting $\nu$ depends on the stability index $\alpha$. Stationary  states exist for $\nu>2-\alpha$, see Ref.~\onlinecite{dybiec2010d}.
For some potential wells exact formulas for stationary states are known.
For $\nu=2$ the stationary state reproduces the noise pulses distribution, i.e. it is given by the $\alpha$-stable density with the same stability index $\alpha$ \cite{chechkin2002,chechkin2003,dybiec2007d} as the noise $\zeta(t)$.
For $\nu>2$ stationary states, even in single-well potentials, are no longer unimodal \cite{chechkin2002,chechkin2003,capala2019multimodal}.
For $V(x)=x^4/4$ and $\alpha=1$ (Cauchy noise) the stationary state is given by  \cite{chechkin2002,chechkin2003,chechkin2004,chechkin2006,chechkin2008introduction}
\begin{equation}
    P_{\alpha=1}(x)=\frac{1}{\pi\sigma_0^{\nicefrac{1}{3}}\left[\left(x/\sigma_0^{\nicefrac{1}{3}}\right)^4-\left(x/\sigma_0^{\nicefrac{1}{3}}\right)^2+1\right]}
    \label{eq:stationary-n4}.
\end{equation}
Probability density given by Eq.~(\ref{eq:stationary-n4}) has two maxima at $x=\pm \sigma_0^{\nicefrac{1}{3}}/\sqrt{2}$.
For more general, polynomial, single-well potentials of $x^{\nu}/\nu$ type, with $\nu=4n$ or $\nu=4n+2$ (where $n$ is integer and positive), stationary states are also bimodal  and given by finite series  \cite{dubkov2007}.
For the parabolic addition to the quartic potential
\begin{equation}
    V(x)=\frac{x^4}{4}+a\frac{x^2}{2} \;\;\; \mbox{with}\;\;a>0 ,
\end{equation}
there is a critical value of $a_c=0.794$ such that for $a>a_c$ the stationary state is no-longer bimodal \cite{chechkin2002,chechkin2003,chechkin2004}.

For the system described by the Langevin equation (\ref{eq:langevin}), one can numerically estimate the time dependent probability density of finding a particle in the vicinity of $x$ at time $t$ under the initial condition $x(t_0)=x_0$ as $P(x,t|x_0,t_0)=\langle \delta(x-x(t))\rangle$.
The time evolution of the probability density $P(x,t|x_0,t_0)$  is described by the fractional Smoluchowski-Fokker-Planck equation
 \cite{samorodnitsky1994,podlubny1998,yanovsky2000}
\begin{equation}
 \pd{}{P}{t} = -\pd{}{}{x} V'(x,t)P + \sigma^{\alpha} \pd{\alpha}{P}{|x|}.
 \end{equation}
The fractional operator $\partial^\alpha/\partial |x|^\alpha$ is the fractional Riesz-Weil derivative \cite{podlubny1998,samko1993} which can be  defined via the Fourier transform $
 \mathcal{F}_k\left( \frac{\partial^\alpha f(x)}{\partial |x|^\alpha} \right)=-|k|^\alpha \mathcal{F}_k\left(f(x)\right).$

Equation~(\ref{eq:langevin}) is the large damping limit of the full Langevin equation
\cite{uhlenbeck1930theory,risken1996fokker}
\begin{equation}
\ddot{x}(t)= - \gamma \dot{x}(t) -V'(x) + \zeta (t).
\label{eq:full-langevin}
\end{equation}
Please note that in the case of full dynamics with $\alpha=2$ the noise strength $\sigma$ depends on the damping parameter $\gamma$, as they are linked by the fluctuation dissipation relation \cite{risken1996fokker,sekimoto2010stochastic}.
Contrary to the Gaussian case, for $\alpha<2$, damping and strength of fluctuations are two independent parameters.
The time evolution of the full probability density associated with Eq.~(\ref{eq:full-langevin}) is described by the fractional Kramers equation.
The joint probability density $P=P(x,v,t|x_0,v_0,t_0)$ evolves  according to the fractional  Kramers equation \cite{risken1996fokker,lutz2001fractional}
\begin{equation}
 \frac{\partial P }{\partial t}=\left[- v \frac{\partial}{\partial x} +\frac{\partial}{\partial v}\left( \gamma v + V'(x) \right) + \sigma^{\alpha} \frac{\partial^\alpha}{\partial |v|^\alpha} \right]P.
 \label{eq:kk}
\end{equation}
\normalsize
Stationary states for the model described by Eq.~(\ref{eq:full-langevin}) and associated with the diffusion equation~(\ref{eq:kk}) exist for every value of the stability index $\alpha$, under the condition that $V(x)$ grows to infinity fast enough. For example, the parabolic potential is sufficient.
For $\alpha=2$ steady states are given by the Boltzmann-Gibbs distribution under the condition that the potential $V(x)$ satisfies the same constraint as in the overdamped case.
For $V(x)=x^2/2$, the stationary state is given by the 2D $\alpha$-stable density \cite{sokolov2011,zozor2011spectral} with the non-trivial, $\gamma$-dependent, spectral measure \cite{samorodnitsky1994}.

Within the current manuscript, we extend analysis of underdamped, anharmonic stochastic oscillators driven by L\'evy noise.
We focus on the Cauchy noise with the noise strength $\sigma_0=1$, as it allows for easy comparison of limiting cases.
We explore the model in the weak damping limit, and through increasing the damping coefficient, we study how results for overdamped regime are restored.
Thanks to this, we extend our understanding of anomalous underdamped setups, which are less studied that overdamped models.
Our numerical analysis are included in Section Results (Sec.~\ref{sec:results}).
The manuscript is closed with Summary and Conclusions (Sec.~\ref{sec:summary}) and accompanied with the Appendix.

\begin{figure}[!h]
\begin{center}
\includegraphics[width=0.75\columnwidth]{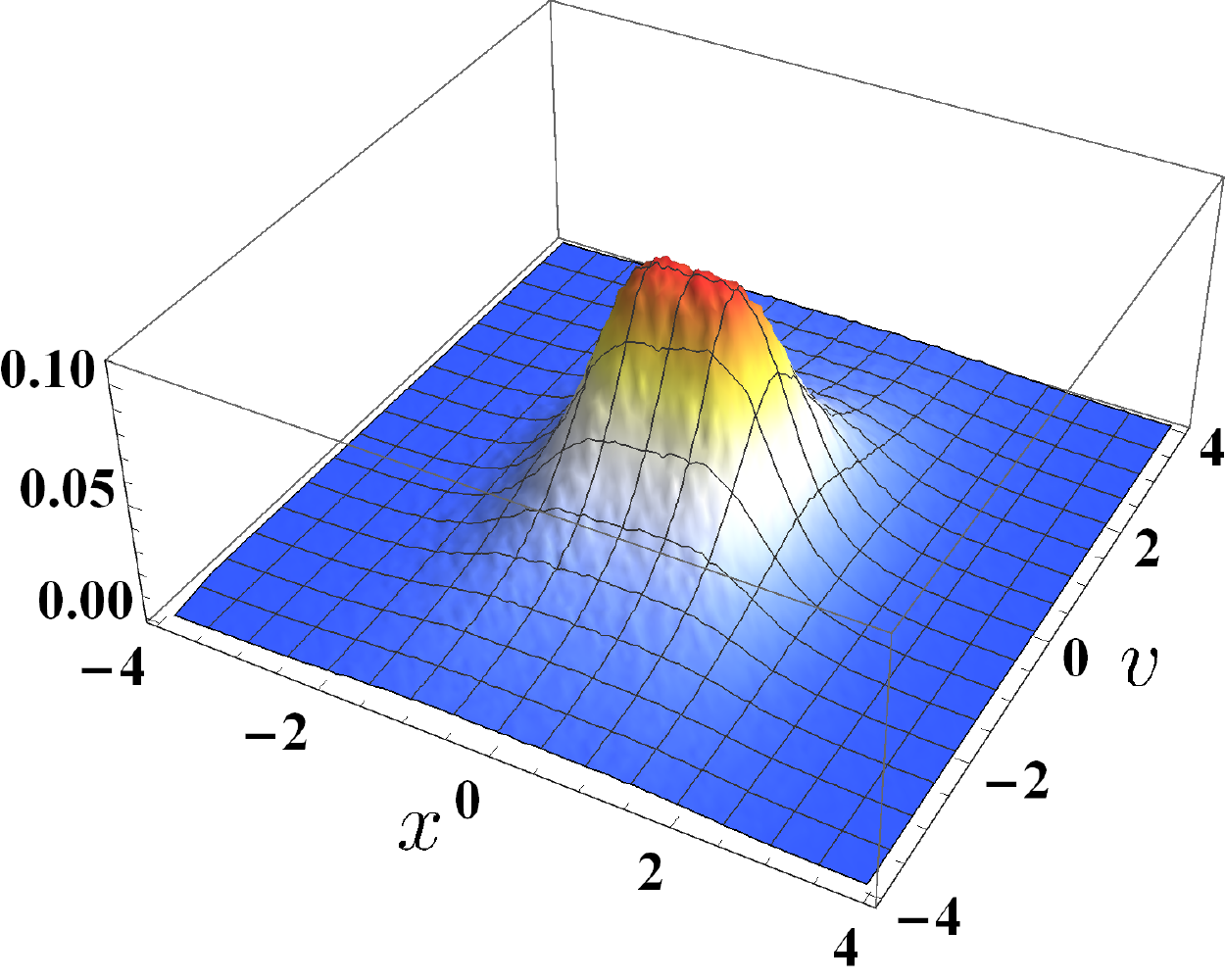} \\ \includegraphics[width=0.75\columnwidth]{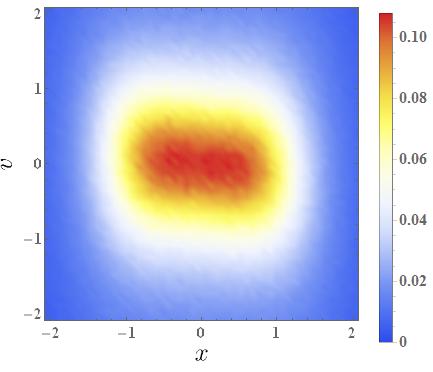}\\
\includegraphics[width=0.75\columnwidth]{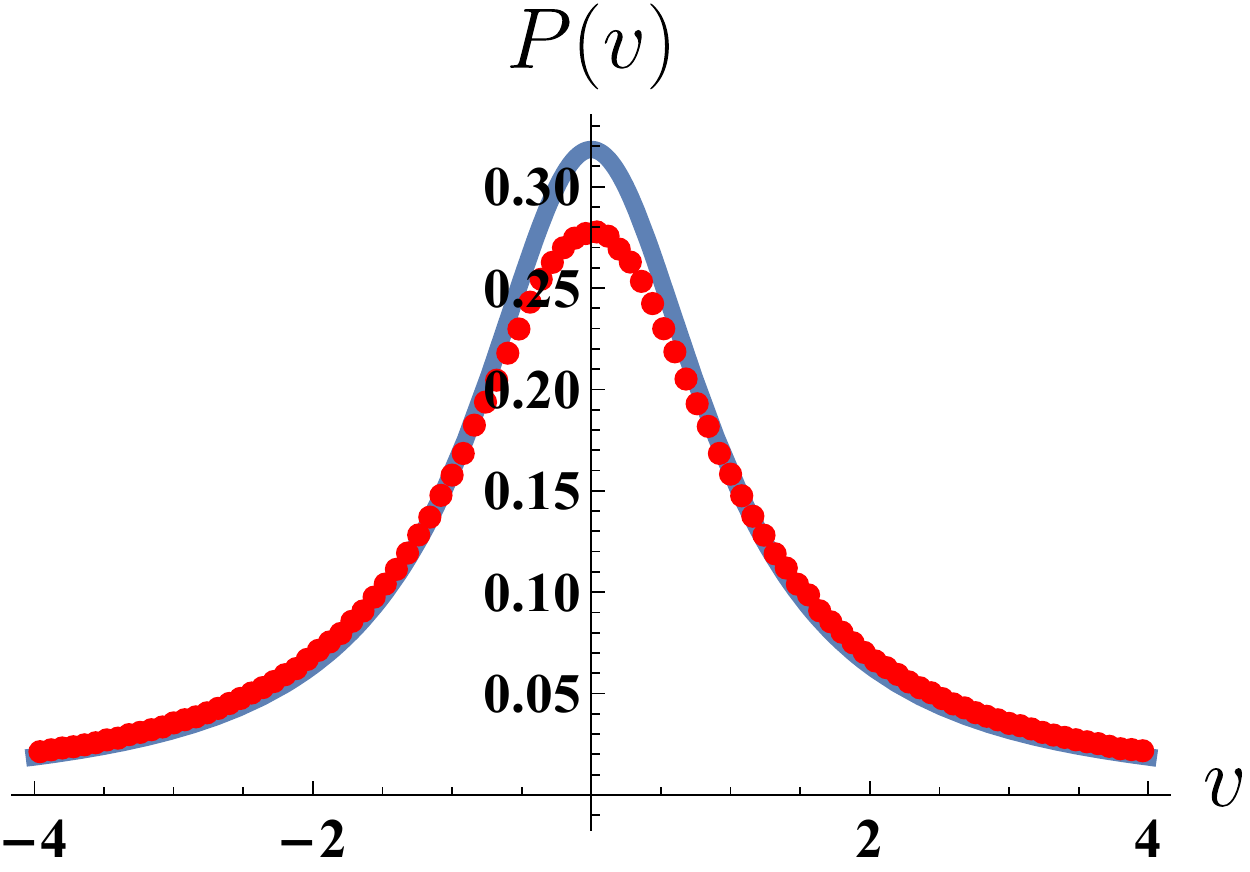} \\ \includegraphics[width=0.75\columnwidth]{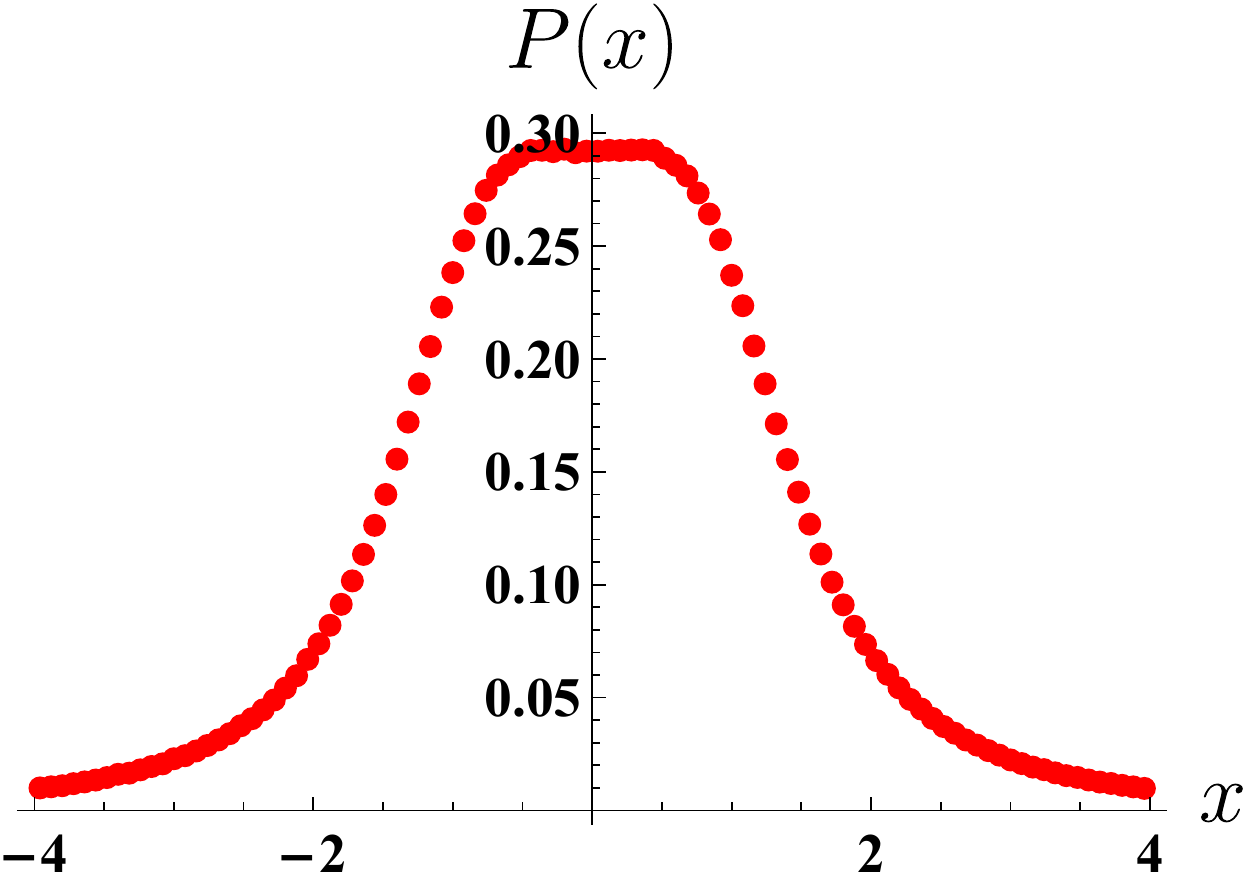}
 \end{center}
 \caption{Stationary probability density $P(x,v)$ as 3D-plot and heat map (top panels), velocity marginal density $P(v)$ (points) with the asymptotic density~(\ref{eq:modpv}) (solid line) and the position marginal distribution $P(x)$ (points) with the asymptotic density~(\ref{eq:stationary-n4}). The damping parameter $\gamma$ is set to $\gamma=1$.
 }
 \label{fig:X4V2g1}
\end{figure}

%%%%%%%%%%%%%%%%%%%%%%%%%%%%%%%%%%%%%%%%%%%%%%%%%%%%%%%%%%
\section{Results\label{sec:results}}

The system described by the full Langevin eqaution~(\ref{eq:full-langevin}) can be studied for any value of the stability index $\alpha$. For $\alpha=2$, the $\alpha$-stable noise is equivalent to the Gaussian white noise. In such a case, for any $V(x)$ such that $V(x)\to\infty$ as $|x|\to\infty$, the stationary state exits and it is given by the Boltzmann-Gibbs distribution
\begin{equation}
    P(x,v) \propto \exp\left[ - \frac{1}{\sigma^2} \left(\frac{v^2}{2} + V(x) \right) \right].
    \label{eq:bg}
\end{equation}
The damping parameter $\gamma$ controls the rate of reaching the stationary state, but it does not affect the shape of stationary state.
Moreover, in the stationary state, see Eq.~(\ref{eq:bg}), despite the functional dependence $\dot{x}=v$, the position and the velocity are statistically independent, because the stationary density factorizes.
The very different situation is observed for $\alpha<2$.
Due to presence of damping, the stationary state exist for a potential $V(x)$ such that $V(x)\to\infty$ as $|x|\to\infty$ fast enough.
For instance, the parabolic potential is sufficient to produce stationary states.
For a fixed potential, the shape of the stationary state depends on the value of $\gamma$. Moreover, in the stationary state, velocity and position are no longer statistically independent. This behaviour is especially well visible for the underdamped stochastic harmonic oscillator \cite{sokolov2011,dybiec2017underdamped}, when stationary states are non-elliptical, 2D $\alpha$-stable densities \cite{samorodnitsky1994} characterized by non-trivial spectral measures \cite{zozor2011spectral}.

\begin{figure}[!h]
\begin{center}
\includegraphics[width=0.75\columnwidth]{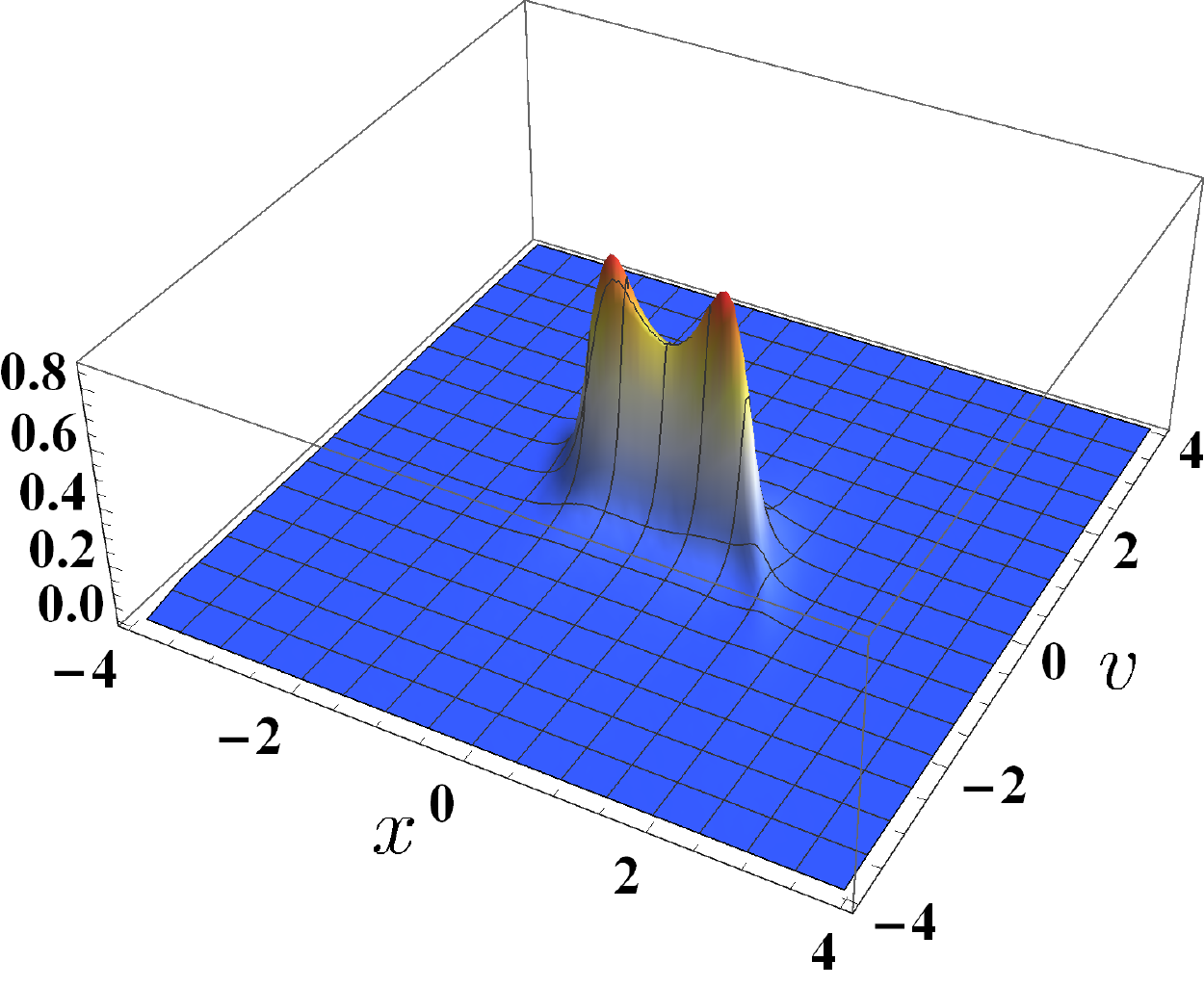} \\ \includegraphics[width=0.75\columnwidth]{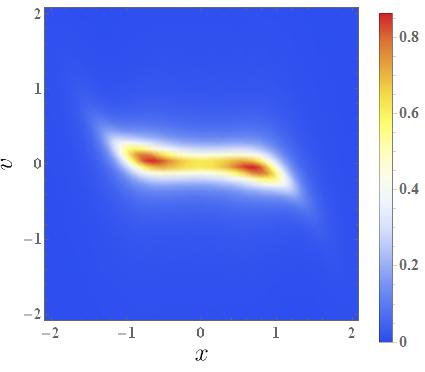}\\
\includegraphics[width=0.75\columnwidth]{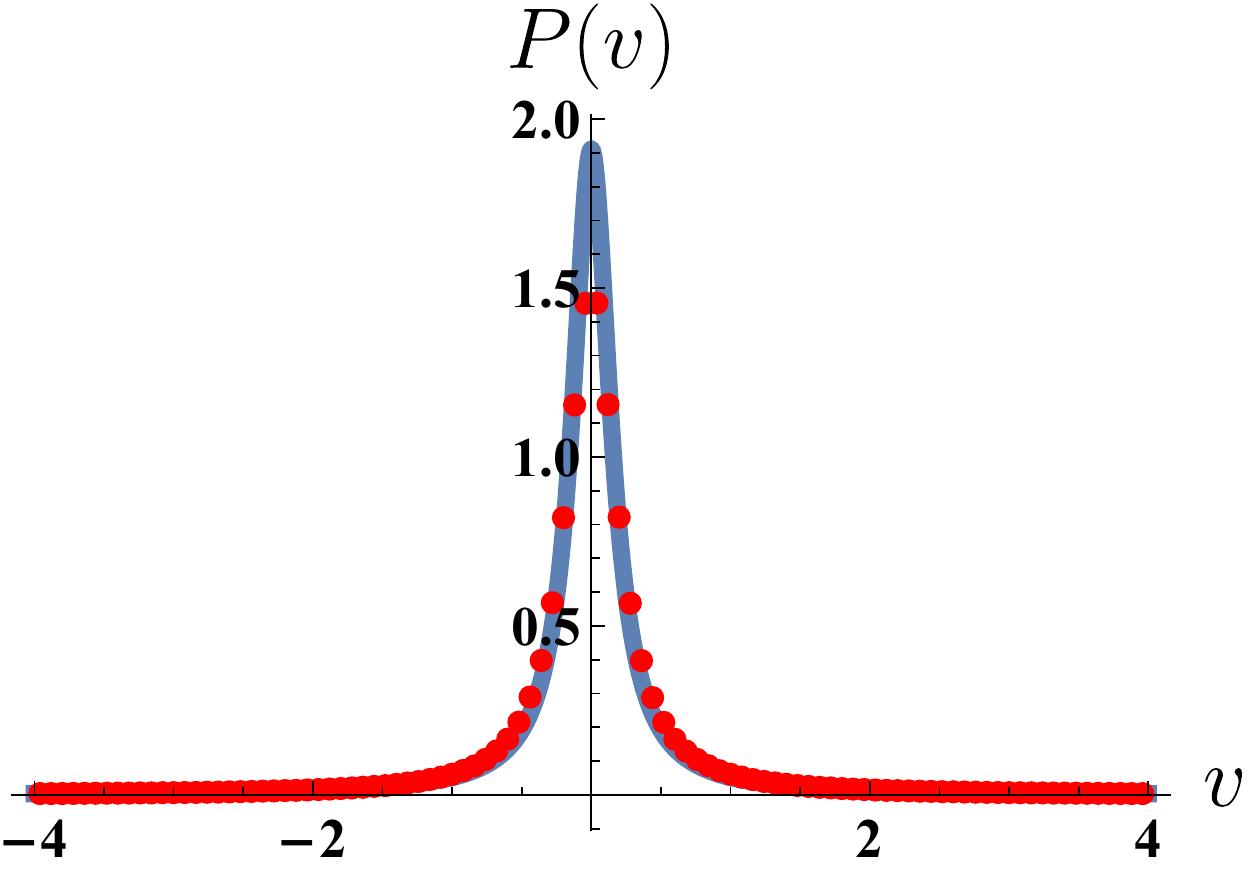} \\ \includegraphics[width=0.75\columnwidth]{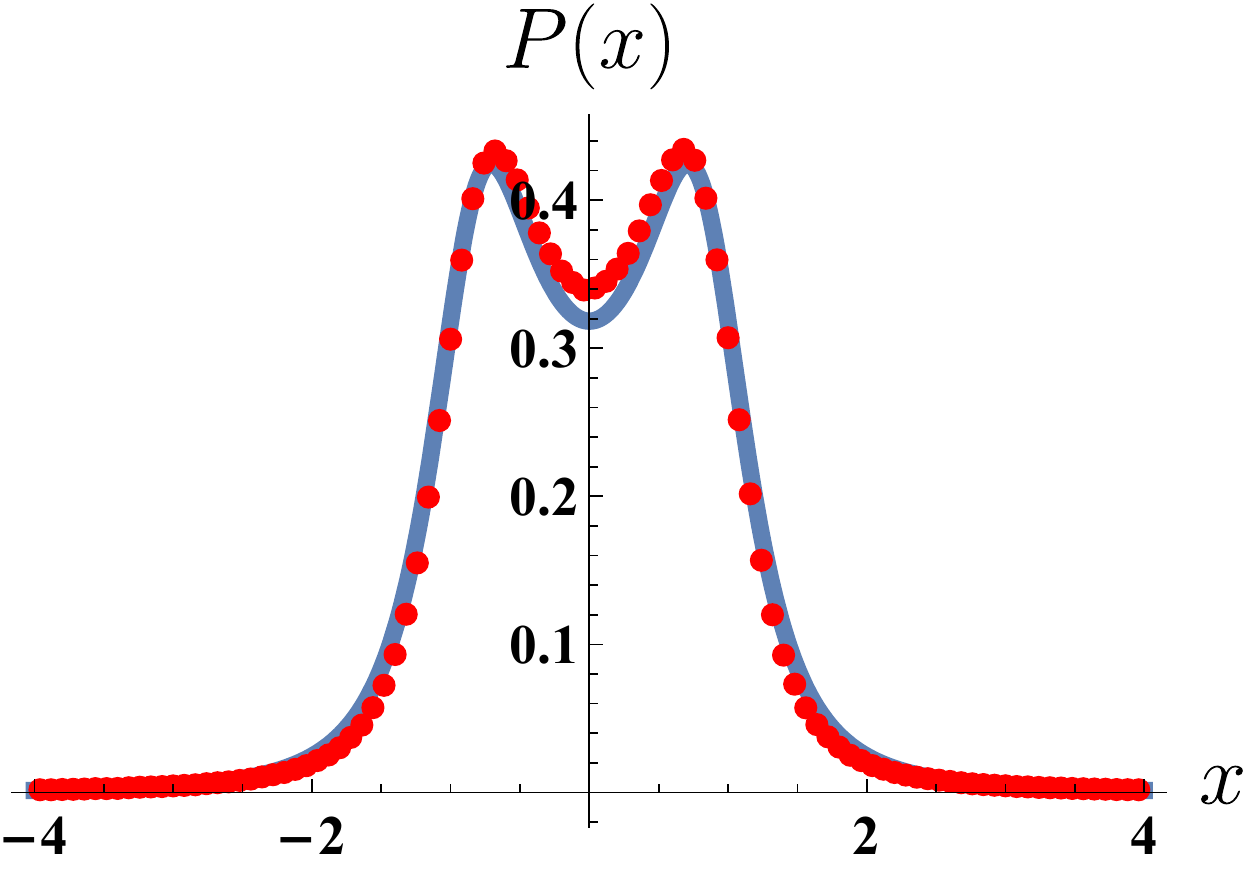}
 \end{center}
 \caption{The same as in  Fig.~\ref{fig:X4V2g1} for $\gamma=6$.}
 \label{fig:X4V2g6}
\end{figure}

Exploration of the full dynamics, allow us to verify under which conditions stationary states reproduce steady states recorded in overdamped systems. 
We restrict ourselves mainly to $\alpha=1$, because for the simplest overdamped systems driven by the Cauchy noise exact results are known. Such a special choice simplifies the comparison of numerical results with known asymptotic regimes.

Results presented in this section have been constructed numerically by simulations of the underdamped Langevin equation~(\ref{eq:linfric}).
% methods of stochastic dynamics.
The part with the $\alpha$-stable noise, e.g. $\dot{v}=-\gamma v -V'(x) + \zeta(t)$, has been integrated with the stochastic Euler-Maryuama method \cite{janicki1994,janicki1996}.  The positions $x(t)$ have been constructed trajectorywise from $v(t)$ realizations.
The Langevin equation~(\ref{eq:full-langevin}) has been integrated with the integration time step $\Delta t=10^{-3}$ and averaged over $N=10^7$ realizations.
From constructed trajectories time dependent and stationary states have been constructed.

The case of $\nu=2$ was explicitly studied in Refs.~\onlinecite{sokolov2011,zozor2011spectral}, where it was shown that the stationary density is given by the 2D $\alpha$-stable density.
Therefore, we start with the stochastic, underdamped, quartic oscillator ($V(x)=x^4/4$)  driven by the Cauchy noise $\zeta(t)$.
The Langevin equation, Eq.~(\ref{eq:full-langevin}), can be rewritten as
\begin{equation}
\left\{
\begin{array}{l}
\dot{v}(t)=-\gamma v - x^3+ \zeta(t) \\
    \dot{x}(t)=v(t)
\end{array}
\right..
\label{eq:linfric}
\end{equation}

One might expect that, analogously like in the overdamped case, the stationary state for Eq.~(\ref{eq:linfric}) could be bimodal also in the underdamped regime.
By exploring dependence of the shape of the stationary state on the damping parameter, we will show that it is not always the case.
On the one hand, in the absence of damping ($\gamma=0$) there is no stationary state for the model described by Eq.~(\ref{eq:linfric}) because the diffusive packet expands boundlessly.
On the other hand, Eq.~(\ref{eq:linfric}) in the strong damping limit $\gamma \rightarrow \infty$ reduces to the well-known problem of the overdamped Cauchy oscillator \cite{chechkin2002,chechkin2003}
\begin{equation}
\dot{x}(t) = -x^3 + \zeta (t).
    \label{eq:langevin_n4}
\end{equation}
As it was mentioned in the introduction, this model has the bimodal stationary state given by Eq.~(\ref{eq:stationary-n4}).
Nevertheless, it is unknown what happens for weak damping (small $\gamma$) and how the transition from weak to strong damping is reflected in the shape of stationary densities.

Figure~\ref{fig:X4V2g1} presents the stationary state for the model described by Eq.~(\ref{eq:linfric}) with $\gamma=1$.
Subsequent panels (from top to bottom) present the 3D surface, 2D heat map and marginal $P(v)$ and $P(x)$ densities.
The stationary density is unimodal and it reflects symmetries of  the potential.
The velocity marginal, $P(v)$, and position marginal, $P(x)$, densities are unimodal as well.
Solid lines in bottom panels of Fig.~\ref{fig:X4V2g1} present the limiting velocity distribution, see Eq.~(\ref{eq:modpv}), and the stationary state for the overdamped quartic Cauchy oscilator, see Eq.~(\ref{eq:stationary-n4}).
For $\gamma=1$, limiting marginal densities differ from recorded $P(v)$  and $P(x)$ marginal densities.

For the increasing damping $\gamma$, distributions of a velocity become narrower, because the acceleration of the particle becomes hindered.
This is confirmed by Eq.~(\ref{eq:modsigma}), which demonstrates that, with the increasing $\gamma$, the scale parameter $\sigma$ is reduced.
At the same time, the impact of rare `long jumps' in velocity becomes limited.
This is reflected in the shape of position marginal distribution, which becomes similar to the solution of the overdamped quartic Cauchy oscillator.
The increase in  $\gamma$ leads eventually to appearance of bimodality in $P(x)$ at $\gamma \approx 1.5$.
Maxima of the full probability density still are placed  at $v=0$ but
$x=0$ becomes saddle point (maximum in the velocity distribution and local minimum in the position density).
Just above $\gamma=1.5$ maxima are located near $x=0$ and their height is practically negligible.
Nevertheless, the further increase in the damping parameter makes them more pronounced.
With increasing $\gamma$, maxima moves towards larger $|x|$ and become well separated from the background.
Finally, for even larger $\gamma$ ($\gamma \gg 1.5$) the damping starts to play the dominating role in Eq.~(\ref{eq:linfric}), and the process becomes practically overdamped.
This effect is recorded already for $\gamma=6$, see Fig.~\ref{fig:X4V2g6}.
It is well-visible in the velocity marginal distribution, where already for $\gamma=6$ most of probability mass is concentrated around $v=0$.
At the same time, the position marginal distribution is similar to the solution of the overdamped Cauchy oscillator, which is given by Eq.~(\ref{eq:stationary-n4}).
Consequently, a particle most likely has zero velocity, $v=0$, and it is most likely localized around $x=\pm1/\sqrt{2}$ which is the position of the maxima of probability density for the overdamped Cauchy oscillator with $\sigma_0=1$.
The bend shape of $P(x,v)$ density, see the second from the top panel of Fig.~\ref{fig:X4V2g6}, is produced by the deterministic force.
For instance, if $x \gg 0$ there is a strong deterministic force towards origin.
This negative restoring force is responsible for the large (negative) value of the velocity.
Analogously, for $x  \ll 0$, the force and the velocity are positive.
Moreover, as demonstrated in bottom panels of Fig.~\ref{fig:X4V2g6}, for large friction coefficient, i.e. $\gamma=6$, limiting marginal densities, see Eqs.~(\ref{eq:modpv}) and~(\ref{eq:stationary-n4}), are similar to observed $P(v)$ and $P(x)$ marginal densities.

The emergence of a multimimodal stationary state is an effect of the combined action of all three (deterministic, damping and random) forces  which are included in Eq.~(\ref{eq:linfric}).
Noise pulses occasionally give a particle significant velocity, allowing it to move far from the potential minimum.
Deterministic force $-x^3$ is the restoring force.
It pulls the particle back to the potential minimum, thus, it is responsible for the particle acceleration.
The larger distance from the origin, the larger acceleration is.
Simultaneously, with the increase in the velocity, the damping increases.
Therefore, the damping and the deterministic force counterbalance.
If the damping coefficient is large enough, the time needed to deterministically slide to the minimum of the potential becomes infinite.
The probability of visiting the origin can be increased due to random pulses.
Nevertheless, during the sliding the stochastic force typically displaces the particle further away before it reaches $x=0$.
For sufficient large $\gamma$ the number of trajectories not reaching $x=0$ becomes larger than the number of trajectories which visited the potential minimum.
This leads to accumulation of the probability mass outside the potential minimum and emergence of two modal values.

The transition between unimodal and bimodal stationary state, induced by the increase in the damping coefficient $\gamma$, can be also explained in terms of velocity marginal distribution.
Due to ``heavy tails'' of the velocity distribution, ``long jumps'' (abrupt changes) in the velocity are observed.
Because of huge value of the deterministic force $f(x)=-x^3$ at $|x| \gg 1$, displacements to $|x|>1$ are produced by tails of the velocity distribution.
At the same time, with the increasing $\gamma$, the central ($v\approx 0$) part of the velocity distributions becomes narrow, see Eq.~(\ref{eq:modsigma}).
As a consequence, sudden changes in velocity becomes prevailing making the return to the potential minimum unlikely.
This in turn produce the transfer of the probability mass outside the vicinity of the potential minimum.
In contrast, for small $\gamma$, the $P(v)$ distribution is wide, i.e. a lot of the probability mass is located in vicinity of $v\approx 0$.
A particle makes a lot of short jumps which are not sufficient to move a particle to distant points.
Consequently, observed stationary states are unimodal.
Nevertheless, the limit of vanishing damping requires further studies.

The system described by Eq.~(\ref{eq:linfric}) is characterized by two relaxation times $\tau_v$ and $\tau_x$, see Ref.~\onlinecite{chechkin2000linear}.
The damping coefficient controls the rate of velocity relaxation which is characterized by the relaxation time $\tau_v \propto 1/\gamma$.
At the same time, the relaxation in the $x$ is described by $\tau_x \propto (\gamma L/\sigma)^\alpha$, where $L$ is the typical system size.
Situations considered in Figs.~\ref{fig:X4V2g1} and \ref{fig:X4V2g6} differ not only in shape of stationary states but also in relaxation times.
In Fig.~\ref{fig:X4V2g1} the stationary distribution $P(x)$ is reached before $P(v)$ while in Fig.~\ref{fig:X4V2g6} the velocity relaxation is faster than the spatial relaxation.
Nevertheless, we leave for the further studies the detailed examination of the issue of velocity and spatial relaxations in uderdamped Langevin dynamics in single-well potentials.

\begin{figure}[!h]
\begin{center}
\includegraphics[width=0.75\columnwidth]{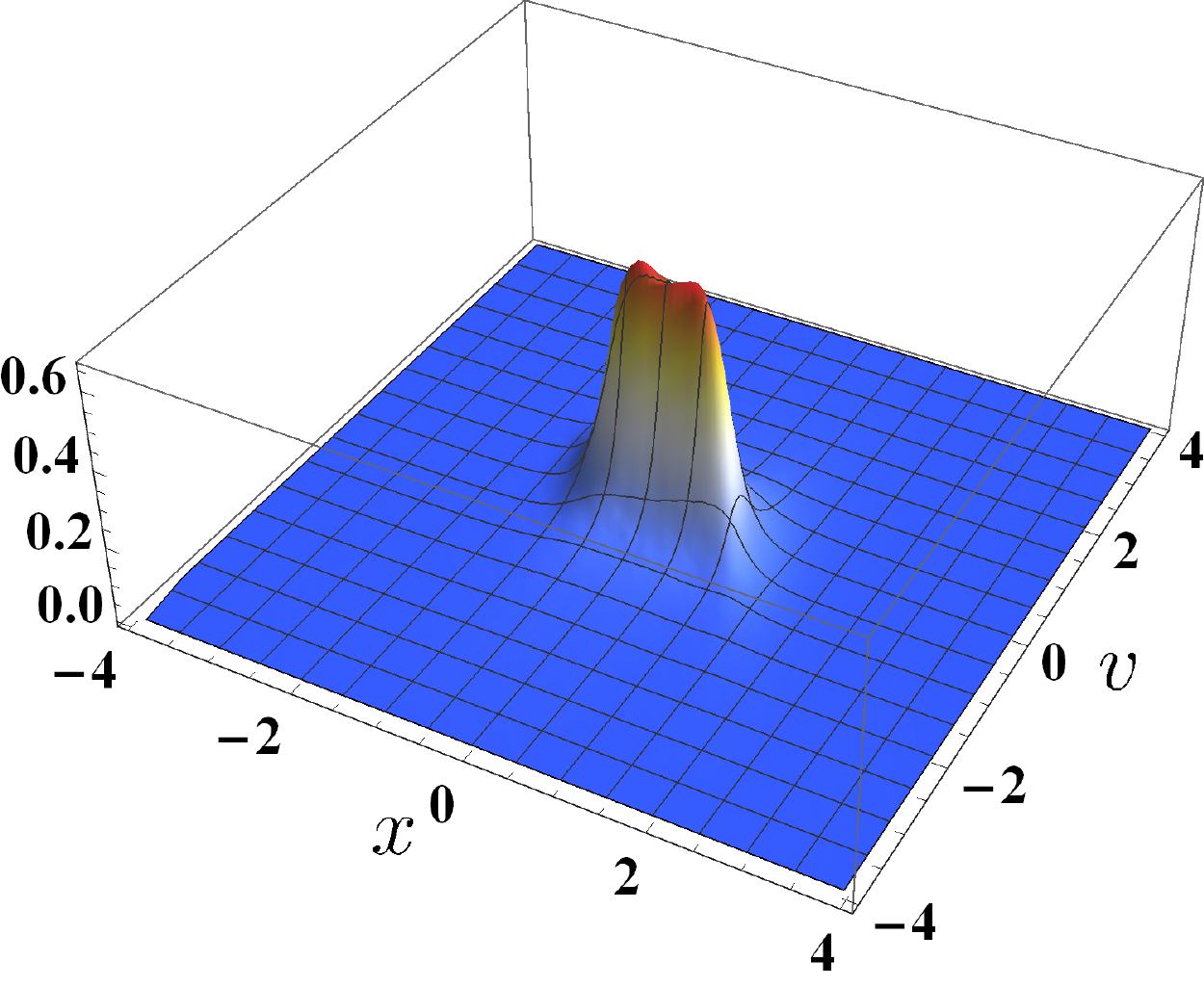} \\ \includegraphics[width=0.75\columnwidth]{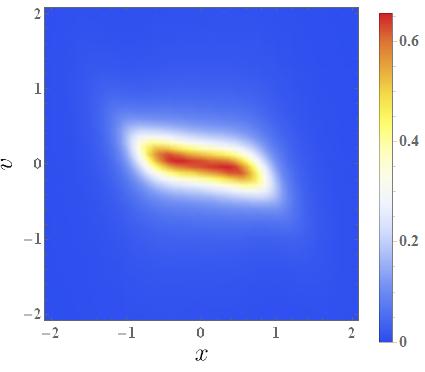}\\
\includegraphics[width=0.75\columnwidth]{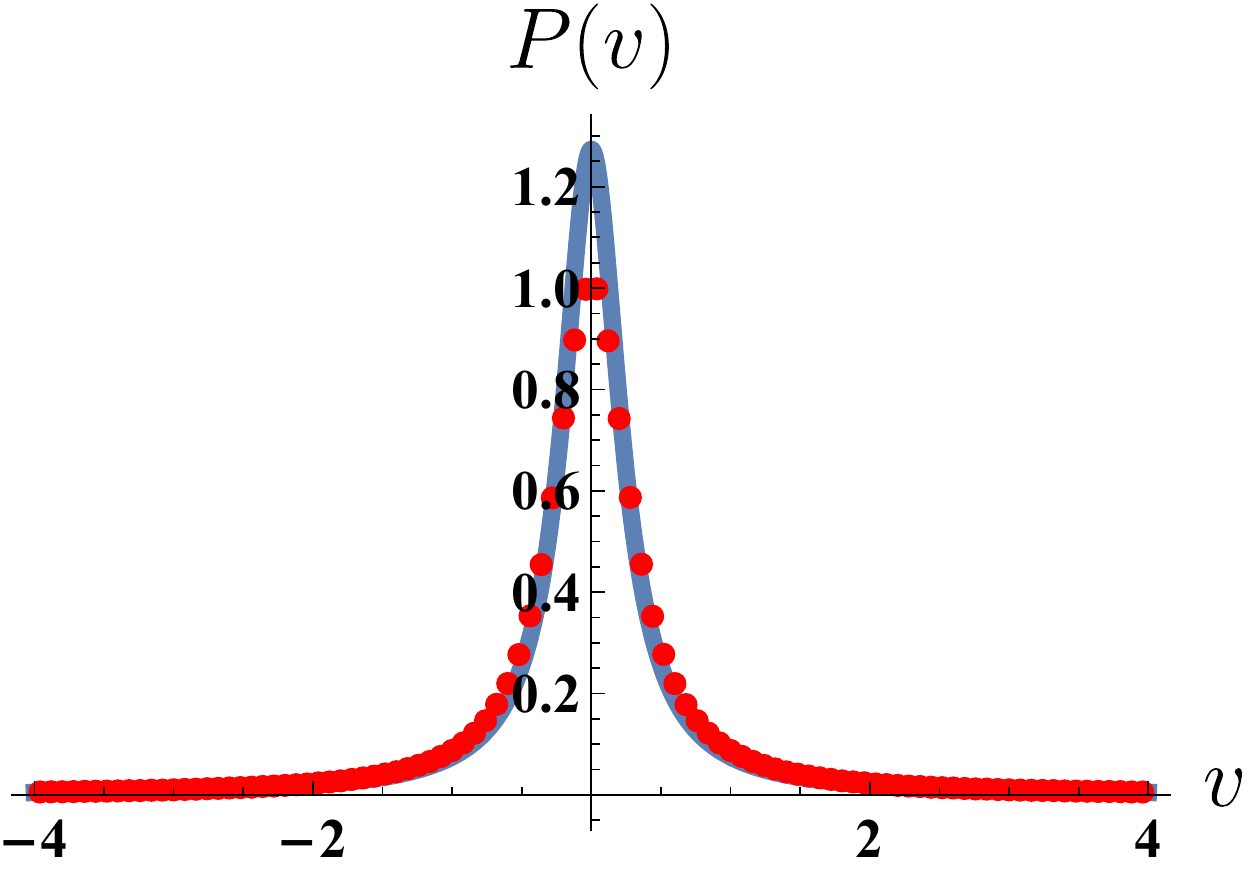} \\ \includegraphics[width=0.75\columnwidth]{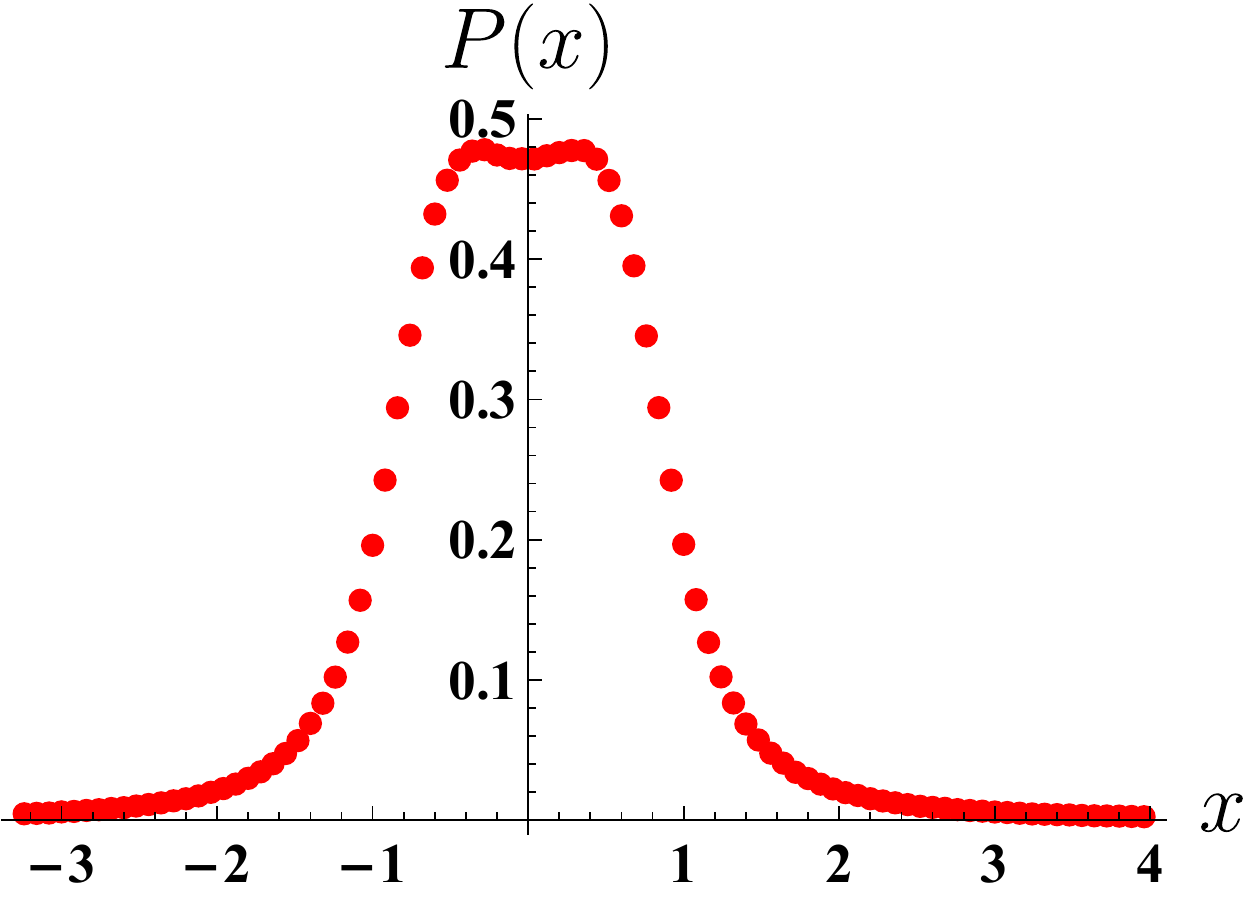}
 \end{center}
 \caption{The stationary state and marginal densities for the potential given by  Eq.~(\ref{eq:apotetial}) with $a=0.5$ and $\gamma=4$.}
 \label{fig:X4V2g4a05}
\end{figure}

The potential used in Eq.~(\ref{eq:linfric}) is the special case of more general binomial potential
\begin{equation}
    V(x)=\frac{x^4}{4}+a\frac{x^2}{2}.
    \label{eq:apotetial}
\end{equation}
In the overdamped regime, for the potential given by Eq.~(\ref{eq:apotetial}), there  exists a critical value $a_c=0.794$, such that stationary state is unimodal for every $a>a_c$, see Ref.~\onlinecite{chechkin2003}.
Consequently, the increase in the strength of the parabolic addition to the quartic potential induce bimodal --- unimodal transition.
In the underdamped regime described by Eq.~(\ref{eq:full-langevin}), one can also explore how results with $a=0$ generalize to $a\neq0$ case.
We expect that for $a>0$ the critical value of the damping parameter $\gamma$ for which the probability function becomes bimodal should increase in comparison to the $a=0$ case, when $\gamma\approx 1.5$.
Indeed, bimodality is observed in simulations with a larger $\gamma$.
For example, for $a=0.5$ a bimodal stationary state is recorded for $\gamma \approx 4$, see Fig.~\ref{fig:X4V2g4a05}.

\begin{figure}[!h]
\begin{center}
\includegraphics[width=0.75\columnwidth]{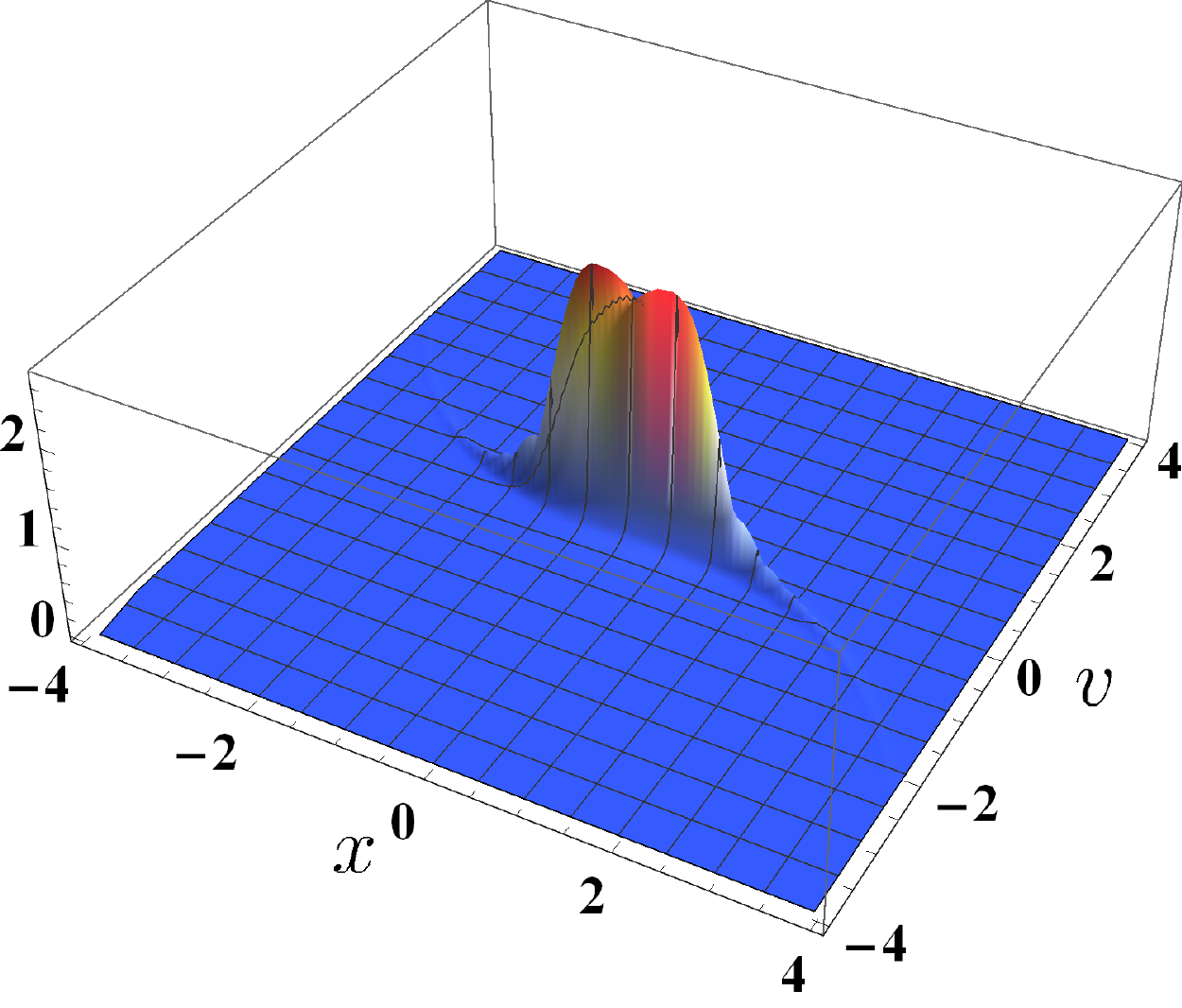} \\ \includegraphics[width=0.75\columnwidth]{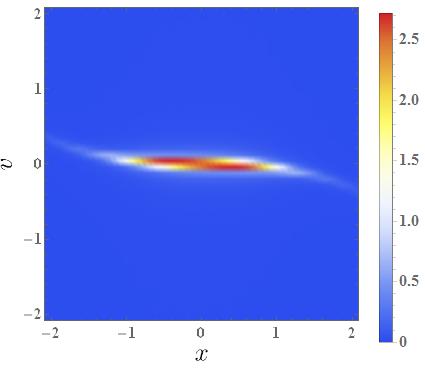}\\
\includegraphics[width=0.75\columnwidth]{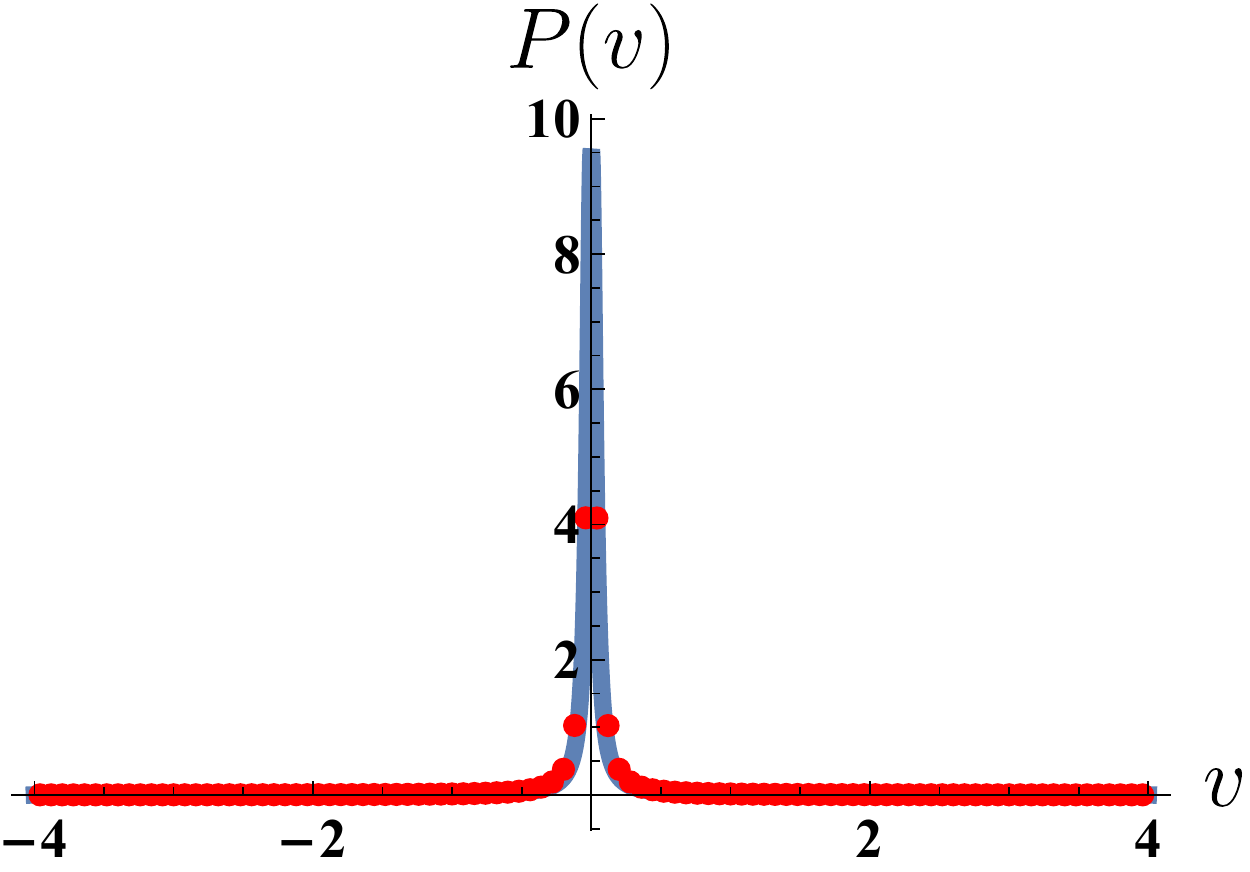} \\ \includegraphics[width=0.75\columnwidth]{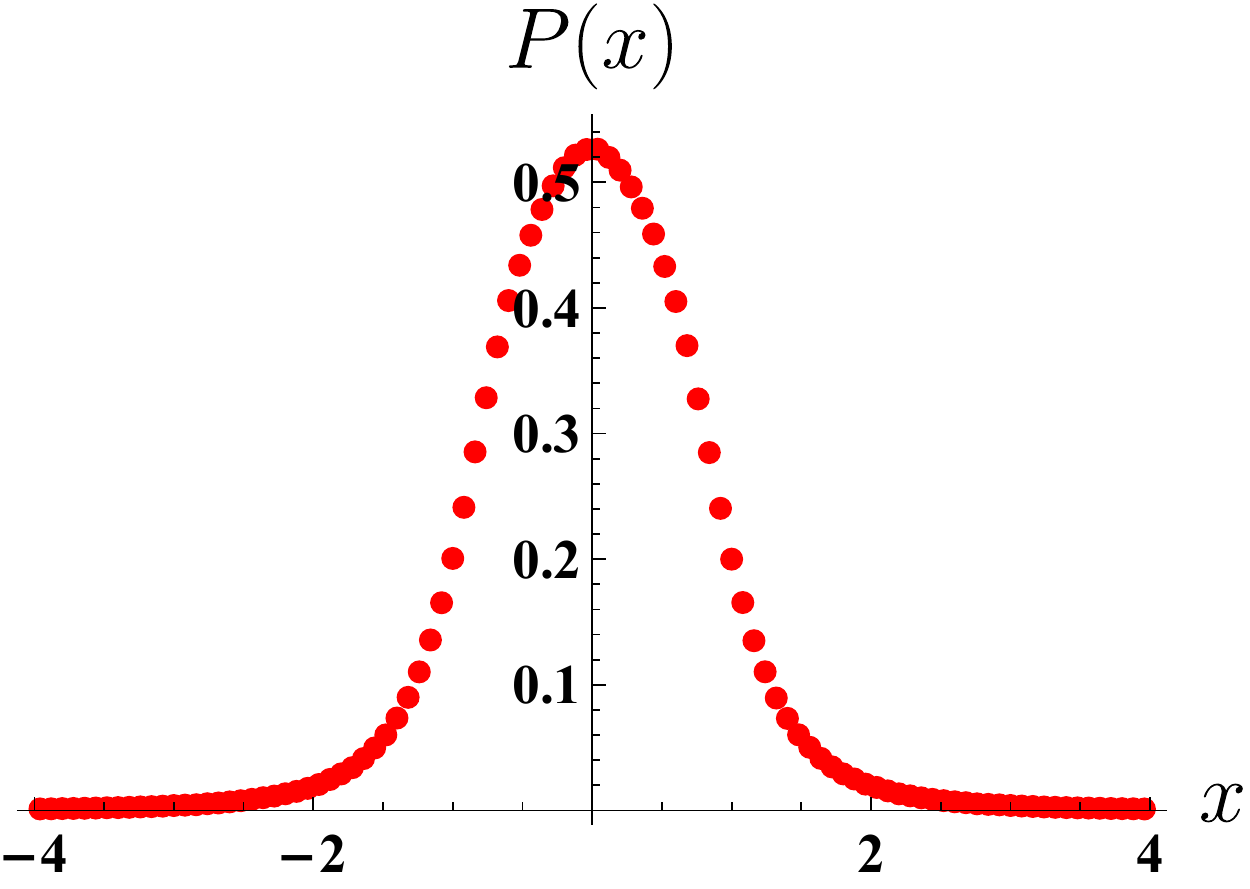}
 \end{center}
 \caption{The same as in  Fig.~\ref{fig:X4V2g4a05} for $a=0.7$ and $\gamma=30$.
 }
 \label{fig:X4V2g30a07}
\end{figure}

A pronounced difference from the $a=0$ case is visible for the parameter $a$ close to the critical value $a_c$.
For instance, for $a=0.7$, the full probability density function has two maxima, see top panel of Fig.~\ref{fig:X4V2g30a07}.
At the same time, position and velocity marginal densities remain unimodal, even for very large $\gamma$, see bottom panel of Fig.~\ref{fig:X4V2g30a07}.

\begin{figure}[!h]
\begin{center}
\includegraphics[width=0.75\columnwidth]{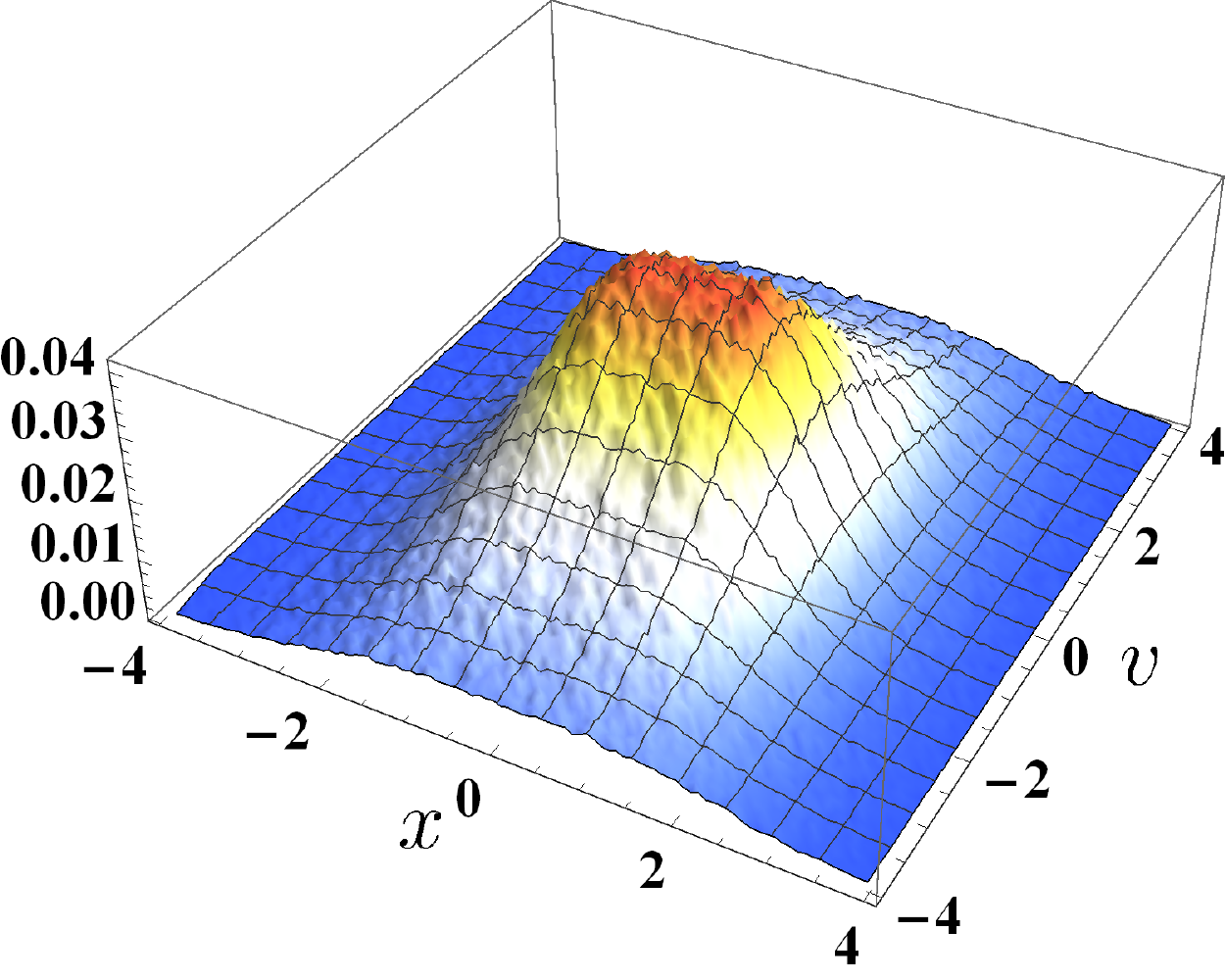} \\ \includegraphics[width=0.75\columnwidth]{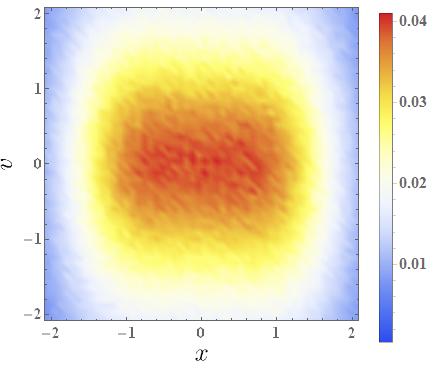}\\
\includegraphics[width=0.75\columnwidth]{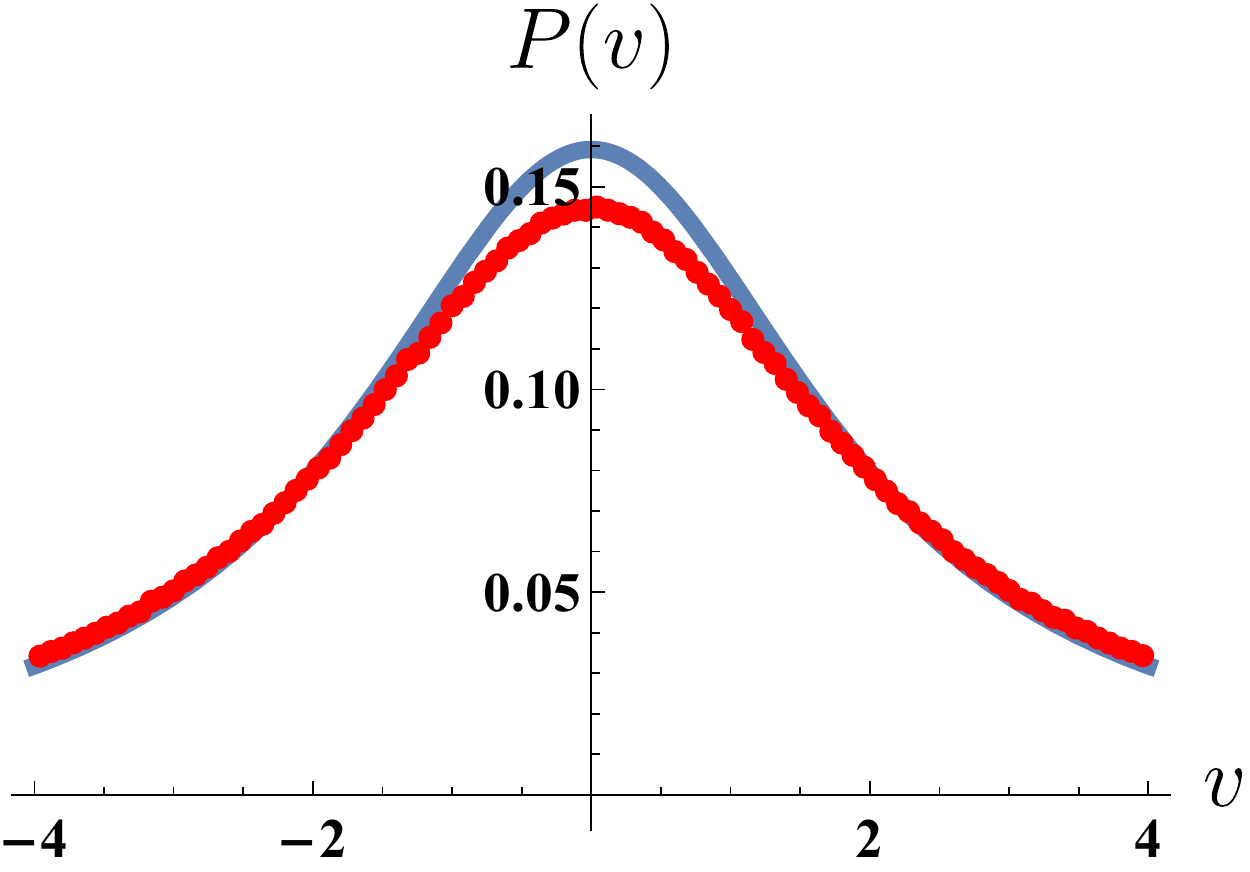} \\ \includegraphics[width=0.75\columnwidth]{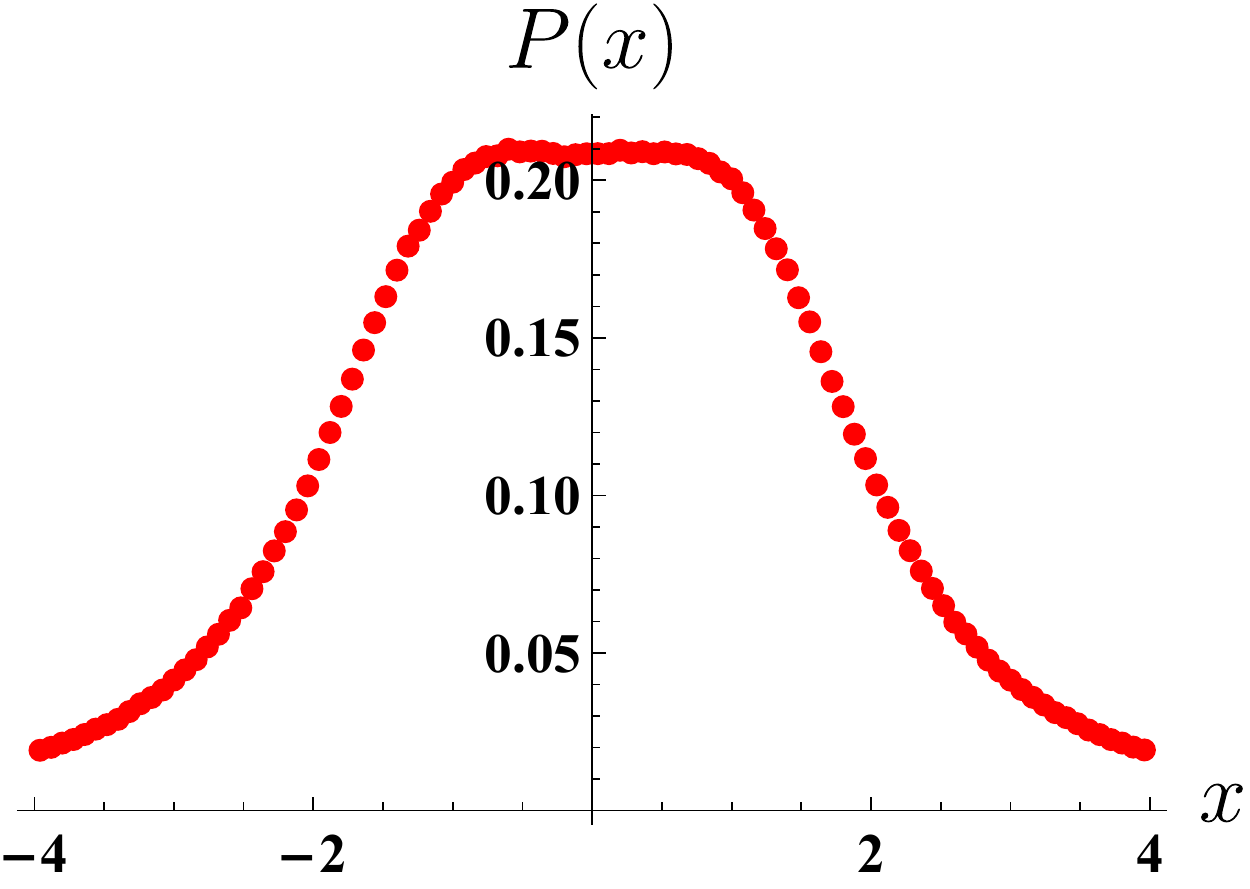}
 \end{center}
 \caption{The same as in  Fig.~\ref{fig:X4V2g4a05} for $a=-0.2$ and $\gamma=0.5$.
 }
 \label{fig:X4V2g05a-02}
\end{figure}

Moreover, for $a<0$,  another counter-intuitive effect is observed.
Due to double-well shape of the potential given by Eq.~(\ref{eq:apotetial}) with $a<0$ one may expect bimodal stationary state.
Nevertheless, if the damping coefficient $\gamma$ is sufficiently small, the probability density seems to be unimodal.
Fig.~\ref{fig:X4V2g05a-02} shows results for $a=-0.2$ and $\gamma=0.5$.
For a small value of the damping parameter $\gamma$, energy is slowly dissipated and the velocity (because of the damping) slowly changes.
Therefore, a particle with a little help from the noise can easily surmount the potential barrier and penetrate neighborhood of both potential minima (unless $a \ll 0$ and $\alpha \lesssim 2$).
When $a$ is further reduced, the stationary state becomes clearly bimodal.

\begin{figure}[!h]
\begin{center}
    \includegraphics[width=0.75\columnwidth]{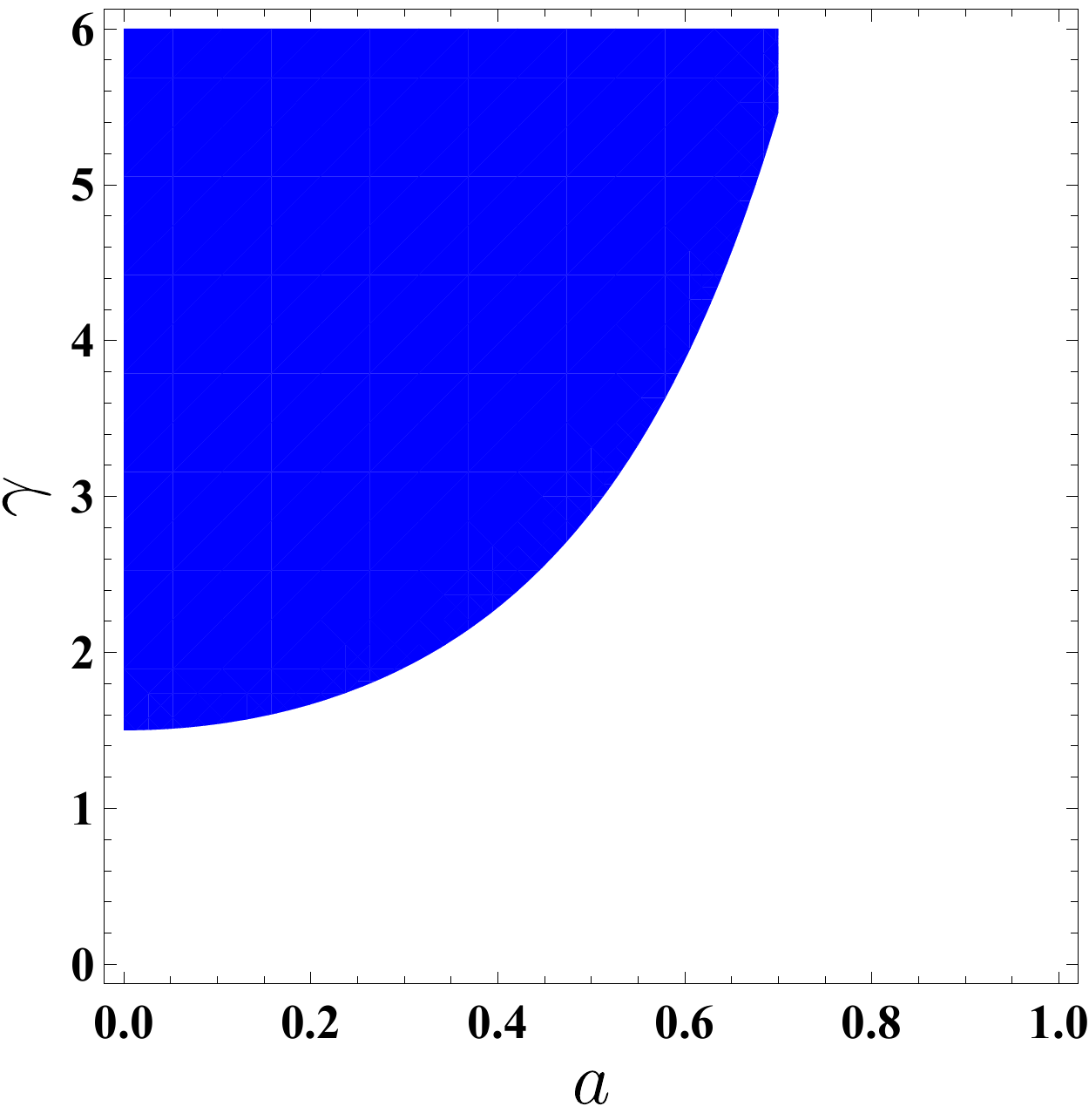}
\end{center}
\caption{The phase diagram for $V(x)=x^4/4+ax^2/2$.
The blue region represents values of parameter $a$ and $\gamma$ for which the full stationary state is bimodal.}
\label{fig:diagram}
\end{figure}

In order to elucidate the role of parameters: $a$ in the potential~(\ref{eq:apotetial}) and the damping $\gamma$ in the full Langevin equation~(\ref{eq:full-langevin}), Fig.~\ref{fig:diagram} presents the phase diagram.
Blue region represents bimodal stationary states, while the white region corresponds to unimodal stationary states.
For $a<0.5$, the value of damping coefficient $\gamma$ for which bimodality appears increases slowly.
For $a>0.5$  the growth of critical damping becomes rapid.
At the same time bimodality of position marginal distribution is not observed.
Finally, for $a>a_c=0.794$, the stationary states are unimodal regardless of the value of the damping parameter.

In the context of the potential~(\ref{eq:apotetial}), it important to discuss the role of the stability index $\alpha$ in more details.
Changes in model properties due to $\alpha$ are milder than due to $a$.
Similarly to the overdamped motion, in the potential given by Eq.~(\ref{eq:linfric}), see Ref.~\onlinecite{chechkin2002}, with the increasing value of $\alpha$, height of peaks decrease and finally they disappear.
At the same time, there is no change in the friction coefficient $\gamma$ for which transition from unimodal to bimodal stationary state occurs or this change is to small to be visible in presented simulations.

\begin{figure}[!h]
\begin{center}
    \includegraphics[width=0.75\columnwidth]{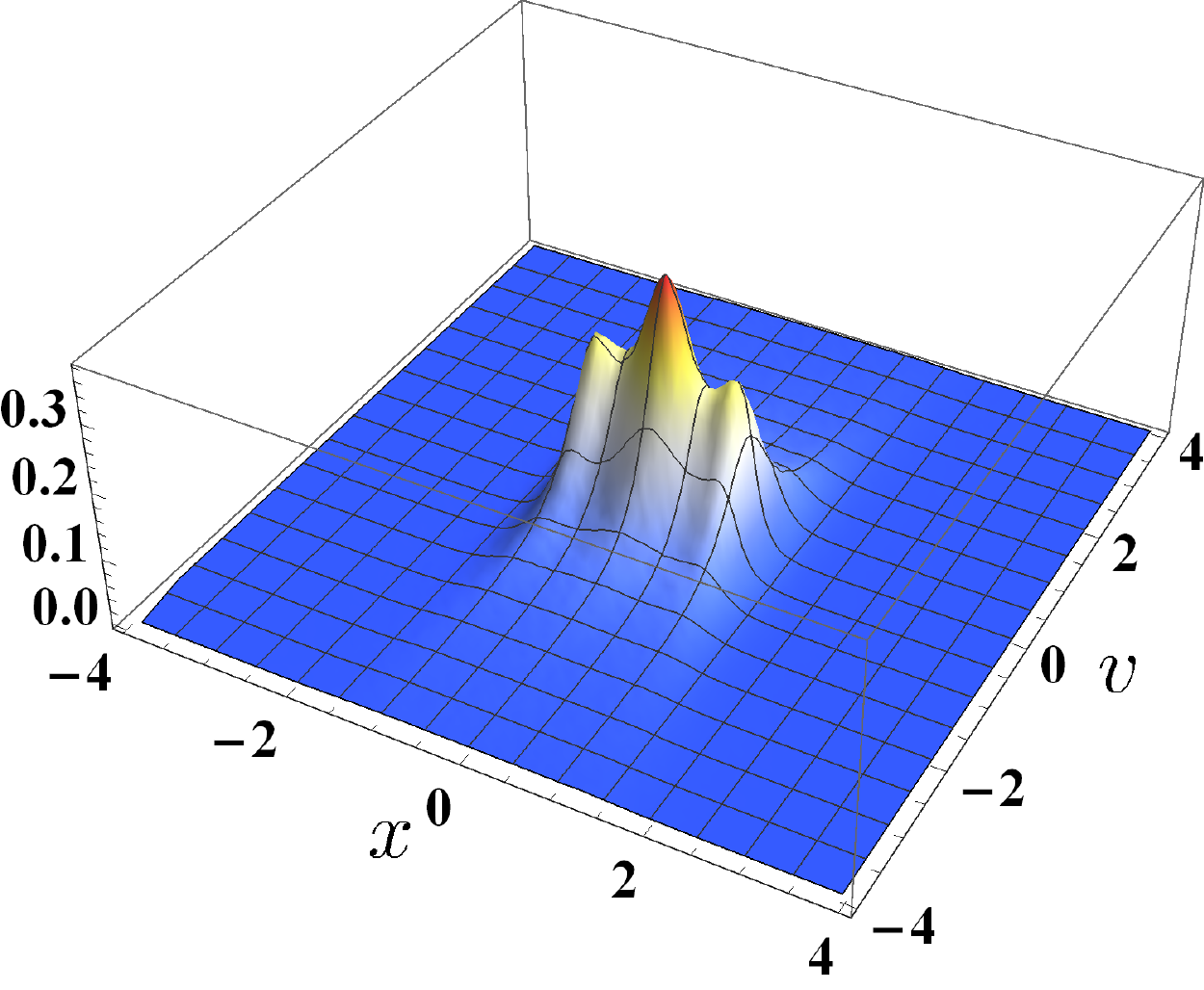} \\ \includegraphics[width=0.75\columnwidth]{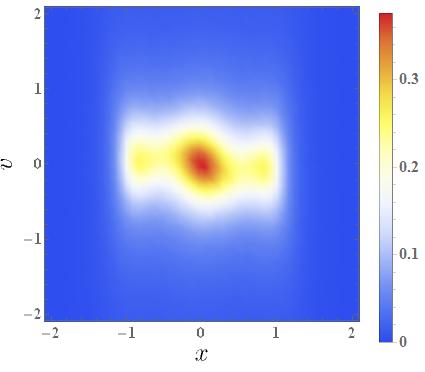}\\
    \includegraphics[width=0.75\columnwidth]{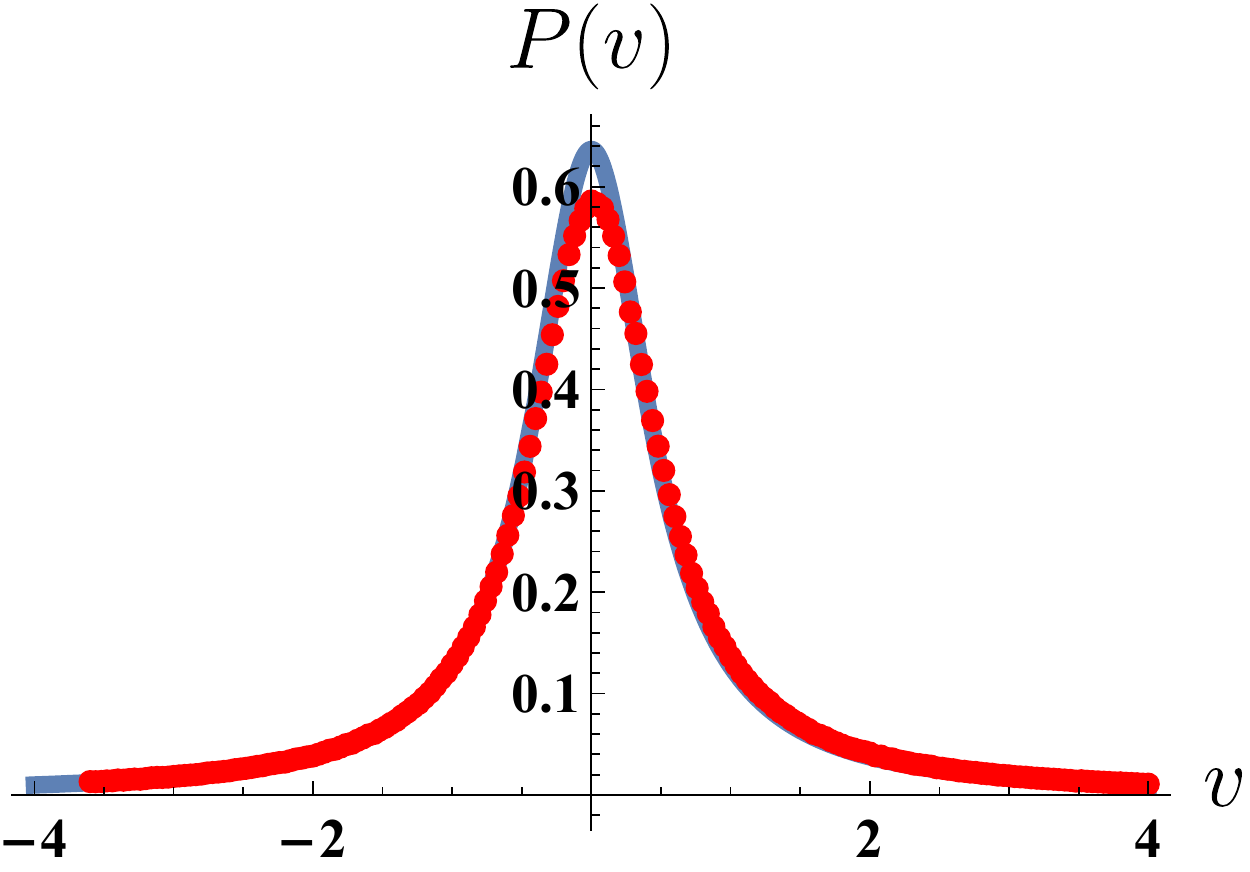} \\ \includegraphics[width=0.75\columnwidth]{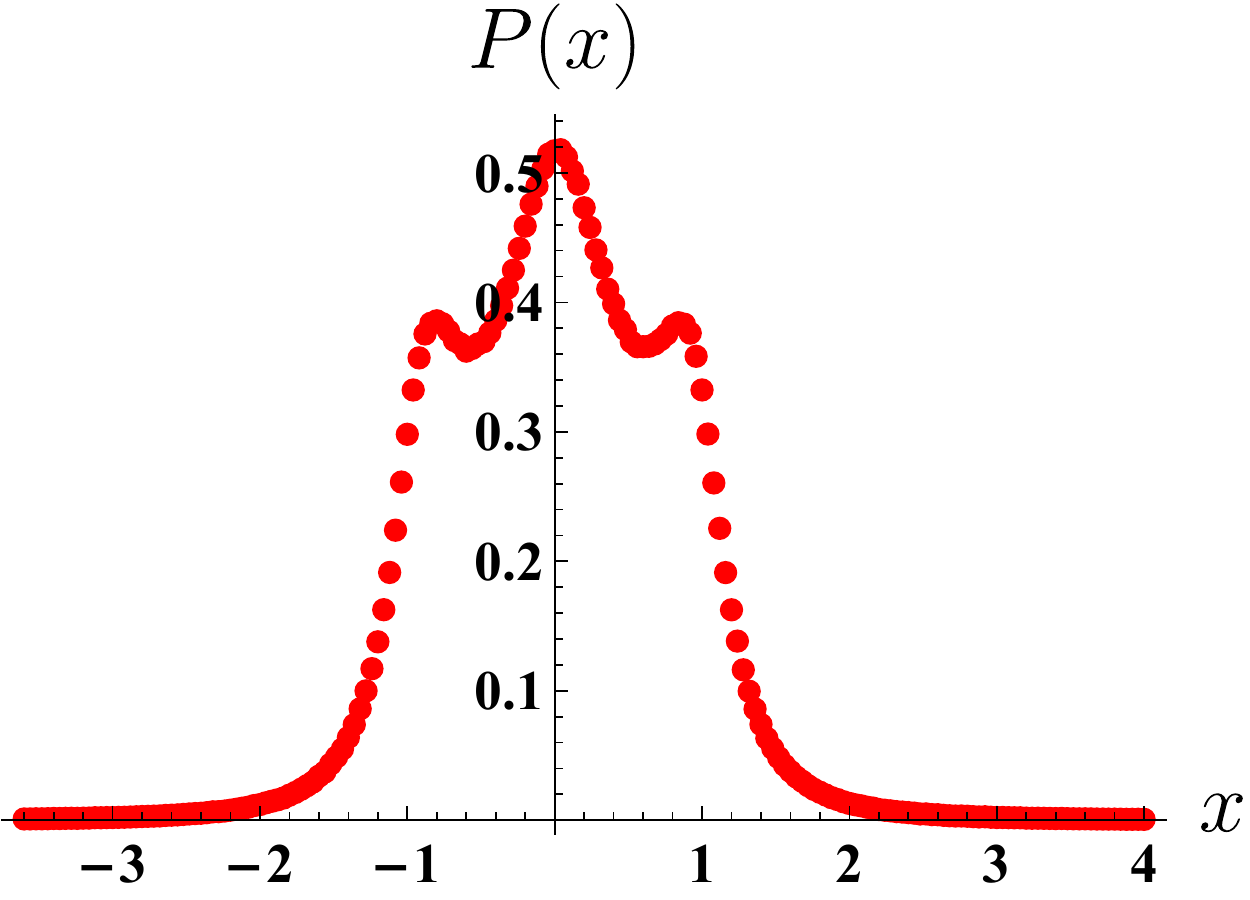}
\end{center}
\caption{
The stationary state and marginal densities for the potential given by  Eq.~(\ref{eq:3mod}) and $\gamma=2$.
}
\label{fig:V2g2mod3}
\end{figure}

We have also examined single-well polynomial potentials.
For instance we have used
\begin{equation}
    V(x)=x^6-\frac{19}{11}x^4+x^2,
    \label{eq:3mod}
\end{equation}
which in the overdamped regime results in the trimodal stationary state \cite{capala2019multimodal}.
For the potential given by Eq.~(\ref{eq:3mod}) behaviour is very similar as for $V(x)=x^4/4$.
For a small value of the damping coefficient $\gamma$, stationary states reflect symmetry of the potential.
When $\gamma$ is large enough additional maxima of stationary density appear.
The exemplary stationary state for the potential given by Eq.~(\ref{eq:3mod}) with $\gamma=2$ is depicted in Fig.~\ref{fig:V2g2mod3}.
With the decreasing $\gamma$, local minima of $P(x)$ at $x\approx \pm 0.6$ becomes shallower. Finally, for sufficiently small $\gamma$ they disappear (results not shown).

Finally, we have explored the motion in the infinite rectangular potential well.
In such a case the motion is restricted in space, because the particle is located within the finite interval.
During interactions with reflecting walls, the velocity changes its sign.
One can expect that, for finite damping, the position marginal density $P(x)$ is uniform while the velocity marginal density $P(v)$ is the same as for a free particle.
Such a shape of marginal densities is confirmed by the Kramers equation  because the stationary density $P(x,v)=C P(v)$ satisfies Eq.~(\ref{eq:kk}).
The very different situation is observed in the $\gamma\to\infty$ limit.
In such a case the motion becomes overdamped. The particle is characterized by the position only. For $\alpha=2$, $P(x)$ is uniform, while for $\alpha<2$ it is u-shaped with modes  at reflecting boundaries \cite{denisov2008}.
Indeed, for finite damping, results of computer simulation show that $P(x,v)=\frac{1}{2L}P(v)$, where $2L$ is the width of the infinite rectangular potential well and $P(v)$ is the same as for the free particle, i.e. it is given by  the $\alpha$-stable density with scale parameter which grows in time.

%%%%%%%%%%%%%%%%%%%%%%%%%%%%%%%%%%%%%%%%%%%%%%%%%%%%%%%%%%
\section{Summary and Conclusions\label{sec:summary}}

Using numerical methods for the Langevin equation,
we have studied stationary states in the underdamped anharmonic stochastic oscillators.
Analogously like in the overdamped case, stationary states exist for potential wells which are steep enough.
Potential wells which asymptotically grow faster than quadratic are sufficient to produce stationary states.
For the parabolic potential, the stationary state exists as well and it is given by the 2D $\alpha$-stable density.
Within these studies, we have used potentials with dominating terms $x^4$ or $x^6$, which are well above minimal required steepness.
The problem of minimal steepness of the potential which is sufficient to produce stationary states in underdamped stochastic oscillators remains open.
We expect that the condition on $\nu$ in $V(x)=|x|^\nu/\nu$ should be not weaker than for the overdamped case, i.e. $\nu>2-\alpha$.

In order to produce multimodal stationary state, e.g. multimodal position marginal density $P(x)$, system dynamics needs to be close to the overdamped regime, i.e. the marginal density $P(v)$ needs to be narrow.
This mean that the damping coefficient $\gamma$ needs to be large enough.
Multimodal stationary states in a single-well potential emerge, when particles are unlikely to be found in the vicinity of the potential minimum.
If the motion is close to the overdamped (narrow $P(v)$), analogously like in the overdamped stochastic oscillators, for $\nu>2$, time required to deterministically slide to $x=0$ is practically infinite.
The sliding is also interrupted by random jumps, which further decrease chances of reaching the minimum of the potential.
Therefore, for $\nu>2$ and $\gamma$ large enough, stationary states can be multimodal.
For large $\gamma$, the motion practically becomes overdamped -- the velocity marginal density $P(v)$ is characterized by the narrow central part, see Eq.~(\ref{eq:modsigma}), while it still has power-law tails.
In the limit of $\gamma\to\infty$ position marginal densities $P(x)$ reproduce those one of overdamped systems.
Appreciable, this equivalence is recorded for finite $\gamma$.
Therefore, for appropriately selected potentials \cite{capala2019multimodal} stationary densities can be characterized by more than two modes.
Consequently, it is possible to fine-tune the potential to produce any given number of modes.

%%%%%%%%%%%%%%%%%%%%%%%%%%%%%%%%%%%%%%%%%%%%%%%%%%%%%%%%%%
\section*{Acknowledgement}

This project was supported by the National Science Center (Poland) grant 2018/31/N/ST2/00598.
This research was supported in part by PLGrid Infrastructure.
Fruitful suggestions from Maciej Majka are greatly acknowledged.

%%%%%%%%%%%%%%%%%%%%%%%%%%%
\appendix
\section{Stationary states in the parabolic potential\label{sec:stationary-parabolic}}

In the case of the linear friction, the evolution of the velocity is described by the following Langevin equation
\begin{equation}
    \frac{dv}{dt}= -\gamma v -V'(x) + \sigma_0 \zeta(t),
    \label{eq:velocity}
\end{equation}
where $\zeta(t)$ is the $\alpha$-stable noise.
Disregarding $-V'(x)$ in Eq.~(\ref{eq:velocity}) results in the linear Langevin equation which is associated with the following Smoluchowski-Fokker-Planck equation
\begin{equation}
    \frac{\partial P (v,t) }{\partial t} = \frac{\partial  }{\partial v} \left[ \gamma v P(v,t)  \right] + \sigma_0^\alpha \frac{\partial^\alpha P (v,t) }{\partial |v|^\alpha}.
\end{equation}
The stationary state fulfills
\begin{equation}
    0 = \frac{d  }{d v} \left[ \gamma v P(v,t)  \right] + \sigma_0^\alpha \frac{d^\alpha P (v,t) }{d |v|^\alpha}.
    \label{eq:stationary}
\end{equation}
Eq.~(\ref{eq:stationary}) in the Fourier space reads
\begin{equation}
    \gamma k \frac{d \hat{P}(k) }{d k } = -\sigma_0^\alpha |k|^\alpha \hat{P}(k),
\end{equation}
where $\hat{P}(k)$ is the Fourier transform $\hat{P}(k)=\int_{-\infty}^\infty P(v) e^{ikv} dv$.
The characteristic function $\hat{P}(k)$ satisfies
\begin{equation}
    \frac{d \hat{P}(k) }{d k} = -\frac{\sigma_0^\alpha}{\gamma} \sign( k) |k|^{\alpha-1} \hat{P}(k).
    \label{eq:st-eq}
\end{equation}
The solution of Eq.~(\ref{eq:st-eq}) is given by
\begin{equation}
    \hat{P}(k) = \exp\left[ -\frac{\sigma_0^\alpha }{\gamma\alpha}   |k|^\alpha \right],
\end{equation}
which is the characteristic function of the symmetric $\alpha$-stable distribution, see Eq.~(\ref{eq:fcharakt}), with the scale parameter
\begin{equation}
\sigma = \frac{\sigma_0}{(\gamma \alpha)^{1/\alpha}}    .
\label{eq:modsigma}
\end{equation}
Consequently, with the increasing $\gamma$, the stationary distribution becomes narrower.
For instance, for the Cauchy noise ($\alpha=1$), the stationary density is the Cauchy distribution
\begin{equation}
    P(v) = \frac{1}{\pi} \frac{\sigma}{\sigma^2+v^2}.
    \label{eq:modpv}
\end{equation}

\end{document}